\documentclass[sigconf, nonacm]{acmart}
\AtBeginDocument{%
  }

\usepackage{graphicx} % Required for inserting images
\usepackage{booktabs}
\usepackage{natbib}
\usepackage{url}
\usepackage{tcolorbox}
\usepackage{fontspec}
\newfontfamily\devanagarifont[Path=fonts/]{NotoSerifDevanagari-Regular.ttf}
\newfontfamily\telugufont[Path=fonts/]{NotoSerifTelugu-Regular.ttf}

\usepackage{multirow}
\usepackage{tabularx}
\usepackage{makecell}
\begin{document}

%%
%% The "title" command has an optional parameter,
%% allowing the author to define a "short title" to be used in page headers.
\title{Building a Cultural Perspective on Doctor-Patient Conversations}

\author{
  \textbf{Krithi Shailya}$^{*1}$,
  \textbf{Siddharth D Jaiswal}$^{*2}$,
  \textbf{Ashish Makani}$^{2}$,
  \textbf{Suvrankar Datta}$^{2}$, \\
  \textbf{Sunayana Sitaram}$^{3}$,
  \textbf{Mohit Jain}$^{3}$
   \\[0.5em]
  $^{1}$IIT Madras, \;
  $^{2}$Ashoka University, \;
  $^{3}$Microsoft Research India
}
\renewcommand{\shortauthors}{}

%%
%% The abstract is a short summary of the work to be presented in the
%% article.
\begin{abstract}
AI-powered medical scribes are increasingly used to transcribe doctor-patient conversations and automate clinical documentation. However, large-scale real-world consultation datasets are scarce due to the sensitivity of clinical conversations, leading developers to rely on simulated and LLM-generated synthetic consultations. While scalable, these alternatives may fail to capture culturally situated patterns of clinical interaction. We introduce interactional cultural markers, measurable patterns of doctor-patient interaction grounded in cross-cultural clinical communication, and use them to compare real, simulated, and synthetic consultations from Indian and US clinical contexts. We find distinct patterns of participation and control: Indian consultations involve greater patient participation but stronger doctor control, while US consultations exhibit balanced participation and open-ended discussion. Synthetic Indian consultations often fail to reproduce these patterns, instead converging toward US-like interaction. We identify additional synthetic signatures, including excessive doctor explanation and formulaic patient responses. We conclude by discussing implications for generating culturally grounded synthetic clinical conversations.

\end{abstract}

%%
%% The code below is generated by the tool at http://dl.acm.org/ccs.cfm.
%% Please copy and paste the code instead of the example below.
%%
% \begin{CCSXML}
% <ccs2012>
%  <concept>
%   <concept_id>00000000.0000000.0000000</concept_id>
%   <concept_desc>Do Not Use This Code, Generate the Correct Terms for Your Paper</concept_desc>
%   <concept_significance>500</concept_significance>
%  </concept>
%  <concept>
%   <concept_id>00000000.00000000.00000000</concept_id>
%   <concept_desc>Do Not Use This Code, Generate the Correct Terms for Your Paper</concept_desc>
%   <concept_significance>300</concept_significance>
%  </concept>
%  <concept>
%   <concept_id>00000000.00000000.00000000</concept_id>
%   <concept_desc>Do Not Use This Code, Generate the Correct Terms for Your Paper</concept_desc>
%   <concept_significance>100</concept_significance>
%  </concept>
%  <concept>
%   <concept_id>00000000.00000000.00000000</concept_id>
%   <concept_desc>Do Not Use This Code, Generate the Correct Terms for Your Paper</concept_desc>
%   <concept_significance>100</concept_significance>
%  </concept>
% </ccs2012>
% \end{CCSXML}

% \ccsdesc[500]{Do Not Use This Code~Generate the Correct Terms for Your Paper}
% \ccsdesc[300]{Do Not Use This Code~Generate the Correct Terms for Your Paper}
% \ccsdesc{Do Not Use This Code~Generate the Correct Terms for Your Paper}
% \ccsdesc[100]{Do Not Use This Code~Generate the Correct Terms for Your Paper}

%%
%% Keywords. The author(s) should pick words that accurately describe
%% the work being presented. Separate the keywords with commas.
%% A "teaser" image appears between the author and affiliation
%% information and the body of the document, and typically spans the
%% page.

% \received{20 February 2007}
% \received[revised]{12 March 2009}
% \received[accepted]{5 June 2009}

%%
%% This command processes the author and affiliation and title
%% information and builds the first part of the formatted document.
\maketitle
{\renewcommand\thefootnote{}\footnotetext{\textsuperscript{*}Work done while at Microsoft Research India.}}

\section{Introduction}

Doctor-patient conversations are increasingly being transcribed to automate clinical documentation, including the generation of clinical notes (e.g., in SOAP format) and electronic health record (EHR) entries \cite{ALBOKSMATY2025105861, asrevaljoel2025}. AI-powered medical scribes have emerged as a prominent approach to this transformation \cite{tierney2025scribes}, with systems such as Abridge, Microsoft Dragon Copilot, and Suki being increasingly adopted in clinical settings in the US \cite{nuance2023dax, suki2025}. This shift is also gaining momentum in India. Government and regulatory efforts to digitize health records, together with recent advances in generative AI and automatic speech recognition, have enabled organizations such as Eka Care and Augnito to develop and deploy AI-based clinical documentation systems at scale \cite{ekacare2025ekascribe, augnito2025}.

% The growing adoption of AI in healthcare is driven by several converging developments, including the digitization of clinical records, advances in automatic speech recognition, and the emergence of clinically capable large language models (LLMs) \cite{blackley2019speech, hodgson2016risks}. 
However, the development and evaluation of AI clinical scribe (ACS) systems face a fundamental data challenge: large-scale, real-world doctor-patient consultation datasets remain scarce \cite{perspectivekoenecke26a}. Clinical conversations contain sensitive health information, making their de-identification difficult, costly, and often impractical at scale~\cite{saley2024meditod,das2024synthetic}. Existing corpora are also unevenly distributed across geographies, languages, cultural settings, and mode of collection. Researchers and developers therefore increasingly rely on \textit{simulated consultations}, in which health professionals and/or patient actors write or enact clinical encounters typically grounded in de-identified clinical notes \cite{ben-abacha-etal-2023-empirical, yim2023acibench, ekacare2025clinicalnote}, and \textit{synthetic consultations}, in which LLMs generate conversations either from scratch or grounded in patient histories \cite{ekacare2025clinicalnote, indicmeddialog2026, wang-etal-2024-notechat}. These consultation datasets provide scalable alternatives to real-world data, and have been used for training and evaluating language models, including for clinical transcription and note generation~\cite{schlegel2023pulsar, liu-liu-2026-synthetic}.
Yet prior work suggests that simulated and synthetic consultations may fail to capture important cultural nuances of clinical communication \cite{li-etal-2023-two}. Such omissions can affect the fidelity of systems that subsequently transcribe consultations, generate clinical notes, or populate EHRs.
% as inaccurate clinical information may propagate into downstream documentation. 
This motivates a systematic comparison of consultation data sources across cultures to assess whether synthetic data preserve culturally situated patterns of clinical interaction. Such comparisons can inform the development and evaluation of clinical conversational AI across diverse clinical contexts.

\begin{figure*}[t]
    \centering
    \includegraphics[width=0.95\linewidth]{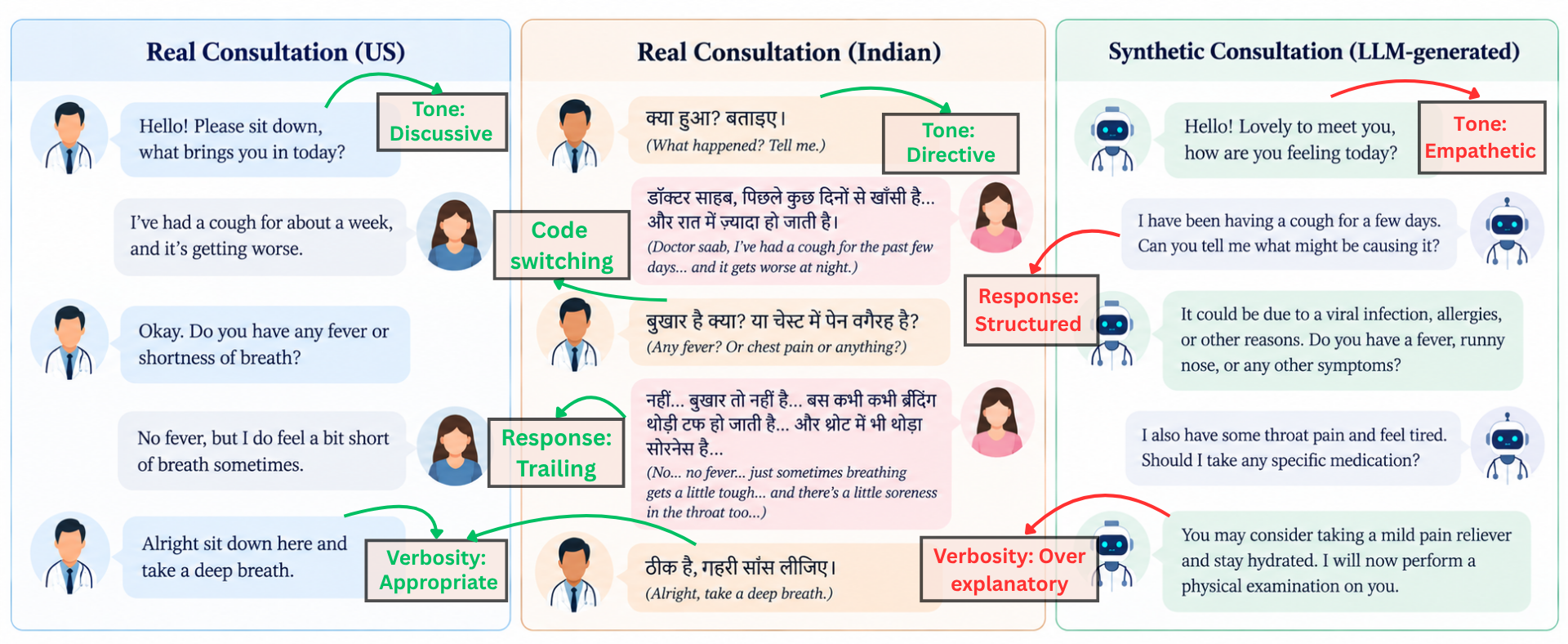}
    \caption{Example conversations across three cultural contexts for the same complaint. The US exchange is more polite; the Indian consultation is directive with respectful, trailing patient responses; the synthetic dialogue diverges from both, with unnatural patient-initiated elicitation, symptom listing, and exam announcements. }
    \label{fig:example_consultation}
\end{figure*}

In this work, we examine how well different consultation data sources capture culturally situated characteristics of doctor-patient interactions. We analyze real-world, simulated, and synthetic consultation datasets from Indian and US clinical contexts, including consultations in English, Hindi, and Telugu. We develop a set of \textbf{interactional cultural markers}: measurable, observable patterns in doctor-patient interaction, grounded in prior work on cross-cultural clinical communication. 
We use these markers to systematically compare consultation datasets, and evaluate synthetic clinical conversations across four generation settings, varying whether generation is grounded in a clinical note and whether it uses a single- or multi-agent architecture. We examine which markers are reproduced, weakened, or introduced through synthetic generation. Figure~\ref{fig:example_consultation} illustrates examples of these cultural markers, highlighting patterns that distinguish real US and Indian consultations from LLM-generated synthetic dialogue. We organize our analysis around three research questions: \textbf{(RQ1)} what interactional characteristics distinguish real Indian clinical consultations from US clinical contexts? \textbf{(RQ2)} To what extent are these characteristics reproduced in synthetic clinical dialogue? \textbf{(RQ3)} How does synthetic cultural fidelity vary across languages, generation settings, and patient characteristics?

We find that Indian and US clinical consultations exhibit distinct patterns of participation and control. Indian consultations involve greater patient participation while retaining stronger doctor control over the progression of the interaction, while US consultations exhibit more balanced participation and more open-ended discussion. Synthetic Indian consultations often fail to reproduce these differences, and instead converge toward more US-like interactional patterns. We also identify strong synthetic signatures that emerge across culture settings, including doctors who over-explain and over-reassure and patients who respond with uniform confidence and formulaic gratitude. Generation strategy introduces further tradeoffs: grounding anchors dialogue to case-specific history-taking but suppresses colloquial language and code-switching, whereas agentic generation produces more naturalistic turn-taking but amplifies verbosity and empathy to unrealistic levels.
These findings demonstrate limitations in using synthetic conversations as proxies for real-world clinical interactions and offer implications for the development and evaluation of culturally grounded synthetic clinical conversational datasets.

\section{Related Work}

\subsection{Clinical Dialogue Data and Synthetic Generation}
Real-world doctor-patient dialogue datasets remain limited. Some of the largest collections are in Chinese, mainly consultation logs from telemedical platforms~\cite{zeng2020meddialog}, but these are text-based asynchronous exchanges rather than recorded in-person consultations. For English, publicly available real-world recorded clinical conversations remain scarce, with existing datasets consisting largely of simulated or synthetic conversations~\cite{ben-abacha-etal-2023-empirical,yim2023acibench, wang-etal-2024-notechat}.
% with annotators writing dialogues grounded in de-identified clinical notes~\cite{ben-abacha-etal-2023-empirical,yim2023acibench}, or a fully `synthetic conversation' generated by prompting LLMs~\cite{wang-etal-2024-notechat}. 
For Indian languages, the situation is starker: to the best of our knowledge, no publicly available real-world multi-turn doctor-patient corpus exists. Existing Indic datasets include simulated conversations \cite{ekacare2025clinicalnote} and task-oriented, code-mixed excerpts \cite{dowlagar2023codemixed}. This scarcity limits the ability to develop and evaluate clinical scribe tools across languages and cultural settings. Building open, culturally situated clinical corpora in Indian and US context remains an important open problem. We hope the framework introduced here provides both the motivation and the tooling to support that effort.

\subsection{Frameworks for Conversational Structure and Culture}
Doctor-patient conversation has long been studied using structured coding frameworks that decompose consultations into observable interactional behaviors. For instance, the Roter Interaction Analysis System (RIAS) categorizes physician and patient utterances across task-oriented and socioemotional dimensions \cite{riasroter}. The Four Habits Coding Scheme focuses on four components of clinician communication: investing in the beginning, eliciting the patient’s perspective, demonstrating empathy, and investing in the end \cite{krupat2006four}. Other frameworks similarly provide structured measures of clinical communication quality and behavior \cite{zill2014measurement}.
These frameworks capture general properties of clinical interaction, but they do not sufficiently capture culturally situated differences in how consultations are organized and conducted, including 
% fine-grained to characterize conversational differences like 
properties such as conversational control, participation structure, and response behavior. 
% As such, they are not directly suited to our goal of assessing whether a conversation reflects the interactional characteristics of a particular cultural setting. But our framework contribution further extends this literature.
Our work builds on this literature of structured interaction analysis and extends it by developing measures suited to examining culturally situated patterns in clinical conversations.

\subsection{Cultural Grounding}
% To define cultural realism in doctor-patient conversations, our goal is to identify conversational characteristics observed in different cultural settings. We review prior studies of doctor-patient communication, focusing on recurring patterns that could be translated into observable cultural properties of a conversation. 
Prior studies of doctor-patient communication provide evidence that clinical interaction differs across cultural settings. Studies of Indian clinical settings report brief consultations, with patients reporting insufficient time for explanation~\cite{waitingtimedas}. Observational audits find doctors relying heavily on closed questions rather than extended discussion~\cite{doctorbehaviour}. Other work characterizes Indian consultations as `directive', with doctors actively guiding the interaction across consultation phases~\cite{mehradoctorinfluence}. Patients typically describe symptoms using everyday idioms and metaphors rather than clinical terminology~\cite{idiomsofdistress}, and conversations frequently involve code-mixing between English and regional languages, including within utterances~\cite{codemixing}.
Studies of US consultations, by contrast, have emphasized greater patient participation, open-ended questioning, discussion, and information exchange~\cite{riasroter}. At the same time, physician-dominant interaction remains common, 
% with biomedically intensive communication
but characterized by greater use of medical terminology~\cite{roterstewart1997, roter2003communication}. Social exchange has also been identified as a structural component of Western clinical consultations, serving as a mechanism for building rapport~\cite{hudakmaynard2011}. Together, these studies qualitatively identify recurring differences in how doctors and patients participate in, organize, and conduct clinical conversations in India and US settings. These observations provide the empirical basis for the interactional cultural markers that we define and quantitatively examine in our work.

\subsection{Towards Culturally Grounded Clinical Dialogue}
Prior work at the intersection of cultural grounding and healthcare has examined how cultural differences shape specific aspects of doctor-patient communication. For example, a qualitative study of Japanese, Korean, and Indian immigrants receiving healthcare in the US found differences in expectations regarding healthcare-seeking, family members' involvement, privacy, disclosure, and end-of-life care~\cite{andresen2001cultural}. Similarly, a cross-national study of 109 physicians in Argentina, Brazil, India, and the US examined how physicians communicated a leukaemia diagnosis, and found that US physicians were more likely to disclose the diagnosis directly and explicitly, whereas physicians in India, Argentina, and Brazil more often disclosed it incrementally or first communicated it to a family member~\cite{torres2007disclosure}. Recent work evaluating LLM-generated doctor-patient conversations used independent clinician ratings of each dialogue across criteria including medical accuracy, realism, persona consistency, empathy, and usability~\cite{haider2025synthetic}.

% \if{0}
% While these studies establish the importance of culture in clinical communication, our understanding of what constitutes cultural variation remains fragmented. Observations about how doctors phrase questions, how patients respond, or whether social exchange occurs at all are scattered across interview-based and qualitative studies, while other work condenses cultural differences into numerical judgment scales. At the same time, many of these interactional properties are not explicitly defined or quantitatively examined in existing dialogue datasets. This makes it difficult to assess whether a dataset actually reflects the clinical environment it is intended to represent. Our work, in comparison builds a structured perspective on cultural clinical dialogue that brings these observations together as measurable properties of interaction. This allows us to empirically analyse how cultural character is expressed, preserved, or lost across real, simulated, and synthetic clinical dialogue.
% \fi
Together, these studies establish that cultural context shapes clinical communication, but they do not provide a systematic way to quantify culturally situated interactional patterns.
The literature remains fragmented: qualitative studies identify recurring behaviors and communication practices, whereas quantitative studies typically assess broad qualities (such as realism, empathy, or usability), often missing the interactional nuances.
% that reduce cultural differences to discrete numerical scales. 
% Neither of these approaches yields quantitative interactional properties that can be used to analyse existing clinical conversational datasets. 
What is missing is a structured set of measurable interactional properties that can be applied consistently across real, simulated, and synthetic consultations.
We address this gap by operationalizing findings from cross-cultural clinical communication as interactional cultural markers, 
% in the literature and consolidate it into a structured set of measurable interactional properties, 
enabling empirical analysis of how culturally situated interactional patterns are expressed, preserved, or lost across consultation data sources.
\section{Methodology}
\label{sec:methodology}

\begin{table*}[]
\caption{Interactional cultural marker annotation framework. \textit{Speaker} indicates which speaker the marker applies to (D-Doctor, P-Patient).
\textsuperscript{\dag}Exam layer is evaluated only when a physical examination is detected.}
\label{tab:framework}
\small
\begin{tabular}{p{0.07\linewidth} p{0.73\linewidth} p{0.08\linewidth}}
\hline
\multicolumn{1}{l}{\textbf{Level}} & \multicolumn{1}{c}{\textbf{Markers}} & \multicolumn{1}{l}{\textbf{Speaker}} \\ \hline

\begin{tabular}[t]{@{}l@{}}
L1\\
\textit{Transcript}
\end{tabular}
&
\begin{tabular}[t]{@{}l@{}}
\textbf{Word Count}: Total words contributed per speaker\\
\textbf{Turn-length Variability}: Variance in turn length\\
\textbf{Backchannel Rate}: Proportion of turns with acknowledgments (eg: \textit{``hmm``, ``okay``})\\
\textbf{Vocative Frequency}: Proportion of turns addressing the doctor (e.g.\ \textit{``sir``/ ``doctor''})
\end{tabular}
&
\begin{tabular}[t]{@{}l@{}}D, P\\D, P\\D, P\\P\end{tabular}
\\ \hline

\begin{tabular}[t]{@{}l@{}}
L2\\
\textit{Inter-turn}
\end{tabular}
&
\begin{tabular}[t]{@{}l@{}}
\textbf{Interruption Rate}: Proportion of turns of overlapping speech\\
\textbf{Echo}: Proportion of turns with repetition of the preceding patient/doctor turn\\
\textbf{Topic Continuity}: Proportion of adjacent turns that stay on-topic\\
\textbf{Question-answer Uptake}: Whether the answer is taken up next turn
\end{tabular}
&
\begin{tabular}[t]{@{}l@{}}D, P\\D, P\\D, P\\D, P\end{tabular}
\\ \hline

\begin{tabular}[t]{@{}l@{}}
L3\\
\textit{Turn}
\end{tabular}
&
\begin{tabular}[t]{@{}l@{}}
\textbf{Tone}: Directive, discussive, or empathetic\\
\textbf{Verbosity}: Terse, appropriate, or over-explanatory dialogue\\
\textbf{Clinical Register}: Type of clinical vocabulary: clinical, colloquial, shorthand\\
\textbf{Question Structure}: Single, multiple related, or multiple unrelated questions\\
% [-4pt]
% \multicolumn{1}{c}{\dotfill}\\[-2pt]
\textbf{Response Style}: How the patient answers: structured, trailing, or deflecting\\
\textbf{Response Reaction}: Accepts, questions, expresses concern, or refuses\\
\textbf{Info Volunteering}: If patient responds exactly or adds additional detail\\
\textbf{Symptom Description}: How symptoms are phrased: clinical, colloquial, or deictic\\
\textbf{Certainty}: How confident or uncertain the answer sounds\\[-4pt]
\multicolumn{1}{c}{\dotfill}\\[-2pt]
\textit{Physical Examination}\textsuperscript{\dag}\\
\textbf{Deictic Reference Density}: Rate of spatial or pointing language per exam turn\\
\textbf{Narration Density}: Proportion of exam turns verbalising the action being performed\\
\textbf{Findings Disclosure}: Proportion of exam turns in which findings are stated aloud\\
\textbf{Compliance}: Cooperation with exam instructions
\end{tabular}
&
\begin{tabular}[t]{@{}l@{}}
D\\D\\D\\D\\
% [-4pt]
% \multicolumn{1}{c}{\dotfill}\\[-2pt]
P\\P\\P\\P\\P\\[-4pt]
\multicolumn{1}{c}{\dotfill}\\[-2pt]\\
D, P\\D\\D\\P
\end{tabular}
\\ \hline

\end{tabular}
\end{table*}

% Detail prompts and metrics in the appendix
We introduce a multi-level framework of interactional cultural markers for analyzing clinical conversations. The framework progresses from \textbf{L1 (transcript level)}, through \textbf{L2 (inter-turn level)}, to \textbf{L3 (turn level)}, allowing the same conversation to be examined from aggregate distributional properties to increasingly local interactional behaviours. L1 primarily consists of automated measures computed directly from the transcript, whereas L2 and L3 use categorical markers assigned through LLM-based annotation (e.g., \textbf{Tone}: directive, discussive, or empathetic). For these markers, an LLM is provided with the relevant turn or turn pair and prompted to assign the category that best characterizes the interaction. We additionally apply a conditional set of \textbf{Physical Examination} markers within L3 when an examination is detected. The marker categories and their operational definitions were derived from prior studies of doctor--patient interaction, translating reported interactional tendencies (e.g., directive physician behaviour) into general categories such as  tone to reduce observation-driven bias and capture broader interactional properties. The framework was refined based on feedback from two medical doctors to assess and improve the clinical relevance of the proposed markers for analyzing doctor-patient interactions. Table~\ref{tab:framework} summarizes the framework and the markers at each level.
% different levels detailed below. 

\subsection{Phase Identification}
\label{sec:phase-identification}
Before applying the multi-level analysis, we identify the phase of each consultation turn using LLM-based annotation. Phase information provides contextual grounding for interpreting interactional markers, allowing the same conversational behavior to be evaluated according to its function within the consultation. For example, examining whether doctors provide terse and directive treatment instructions, as qualitatively identified in prior work~\cite{doctorbehaviour}, requires isolating the treatment phase rather than measuring doctor verbosity across the entire conversation. Similar terseness during history taking can instead indicate excessive interrogation. 

We build our phase structure on the consultation model of \citet{byrne1976doctors}, which identifies five main phases of the medical consultation: (i)~establishing a relationship, (ii)~discovering the reason for attendance, (iii)~conducting verbal and physical examination, (iv)~making a diagnosis, and \textit{(v)~prescribing treatment}. We refine this to better capture interactional patterns relevant to our analysis. We divide the opening of the consultation into \textit{Greeting} and \textit{Social exchange}, distinguishing initial greetings from informal chit-chat that may occur throughout the consultation, as pointed by prior work~\cite{hudakmaynard2011}. 
We further divide the reasons for the visit into \textit{Chief complaint}, \textit{History taking}, and \textit{Physical examination}~\cite{rohr2025whereitdoes}, thereby capturing different participation structures. We retain \textit{Diagnosis} and \textit{Treatment} as separate phases. An additional \textit{Closing} phase is added to examine whether consultations contain a dedicated closing or end abruptly, as suggested by prior work~\cite{waitingtimedas}. Because a turn may serve multiple functions, the model is instructed to assign all applicable phase categories.

\subsection{L1 -- Transcript Level}
At L1, we examine aggregate properties of the transcript that characterize how much and how evenly each speaker contributes. These measures are computed directly from the transcript without LLM-based annotation. \textbf{Word Count} measures the number of words produced by each speaker. \textbf{Turn-length Variability} measures the variance in the number of words across a speaker's turns, computed as the coefficient of variation of turn length.
\textbf{Backchannel Rate} measures the proportion of turns containing acknowledgments such as \textit{``okay''} or \textit{``hmm''}, using a multilingual word list of agreement tokens. \textbf{Vocative Frequency} measures how frequently patients explicitly address the doctor using vocatives such as \textit{``sir''} or \textit{``doctor''}, based on a predefined word list.

\subsection{L2 -- Inter-Turn Dynamics} 
At L2, we examine how speakers respond to one another across turn boundaries. We consider adjacent doctor-patient turns and characterize how a subsequent turn takes up, redirects, or echoes the preceding contribution.
\textbf{Interruption Rate} captures whether a doctor turn cuts off or redirects an ongoing patient narrative, without taking up its unresolved content. The model is given two adjacent turns and assigns a binary label indicating whether the doctor's turn redirects the patient's preceding trailing turn.
\textbf{Echo} captures whether a doctor turn begins by repeating or closely reformulating a word or phrase from the patient's immediately preceding turn. For example, Patient: \textit{``... neck stiffness''} $\rightarrow$ Doctor: \textit{``This neck stiffness... were you doing something at the time?''}). 
% The model is given two adjacent turns and assigns a binary label. 
\textbf{Topic Continuity} measures the proportion of adjacent turns that remain on the same topic, based on whether the subsequent turn continues, elaborates, or responds to the topic introduced in the preceding turn.
\textbf{Question-answer Uptake} captures whether an answer is taken up in the next turn, indicating that the subsequent speaker acknowledges, responds to, or builds on the information provided. 
% The model is given the relevant adjacent turns and assigns a binary label.
Together, these measures characterize how speakers coordinate their contributions and manage the progression of the interaction.

\subsection{L3 -- Turn Level} 
At L3, we examine individual turns to characterize how doctors and patients enact their roles within the consultation. The phase identification allows these markers to be interpreted in the context of the consultation phase in which they occur.
For doctors, \textbf{Tone} categorizes turns as \textit{directive}, \textit{discussive}, or \textit{empathetic}. A directive turn provides an instruction or asks a closed question without offering a rationale (e.g., \textit{``Take this twice a day''}); a discussive turn provides reasoning or frames the exchange collaboratively (e.g., \textit{``I'm giving you this because it reduces inflammation... any questions?''}); an empathetic turn explicitly acknowledges the patient's emotional state (e.g., \textit{``I understand this has been worrying for you''}).
\textbf{Verbosity} categorizes each doctor turn as \textit{terse}, \textit{appropriate}, or \textit{over-explanatory}. Terse turns contain only the information required for the clinical purpose, appropriate turns provide a natural level of contextual explanation, and over-explanatory turns contain more elaboration or repetition than the clinical situation requires.
\textbf{Clinical Register} captures the dominant linguistic register of a turn as \textit{formal clinical}, \textit{colloquial}, or \textit{medical shorthand}. Formal clinical language uses standard medical vocabulary (e.g., \textit{``bilateral pedal oedema''}), colloquial language uses everyday phrasing (e.g., \textit{``lightheaded''}), and medical shorthand uses common clinical abbreviations (e.g., \textit{``BP''}).
\textbf{Question Structure} characterizes how doctors organize information requests within a turn as \textit{single}, \textit{multiple related}, or \textit{multiple unrelated}. A single-question turn contains one information request; a multiple-related-question turn contains several questions addressing the same topic or clinical issue; and a multiple-unrelated-question turn combines questions concerning distinct topics or clinical issues.

For patients, \textbf{Response Style} categorizes response as \textit{structured}, \textit{trailing}, or \textit{deflecting}. A structured response directly answers the doctor's preceding question; a trailing response begins by answering but drifts into unsolicited narrative; and a deflecting response pivots away from the question. 
\textbf{Response Reaction} captures how patients respond to the doctor's preceding thoughts or judgments, with categories of \textit{accepts}, \textit{questions}, \textit{expresses concern}, and \textit{refuses}. An accepting response indicates agreement with, acknowledgment of, or willingness to follow the doctor's contribution (e.g., \textit{``Okay, I understand.''}); a questioning response seeks clarification or challenges the doctor's contribution through a question (e.g., \textit{``Why do I need to take it?''}); an expression of concern communicates worry, fear, discomfort, or hesitation in response to the doctor's contribution (e.g., \textit{``I'm worried about the side effects.''}); and a refusal explicitly declines an instruction, recommendation, examination, or proposed course of action (e.g., \textit{``I don't want to take that medication.''}).
\textbf{Info Volunteering} assigns a binary label: \textit{exact} (the patient provides only the requested information) or \textit{add-on} (the patient volunteers additional relevant information).
\textbf{Symptom Description} examines the linguistic framing used by patients to describe their symptoms as \textit{clinical}, \textit{colloquial}, or \textit{deictic}. A clinical description uses medically specific or relatively formal terminology (e.g., \textit{I have sleep apnea''}); a colloquial description uses everyday language or an informal metaphor to describe the experience (e.g., \textit{``head will burst in pain''}; and a deictic description identifies the symptom through reference to a bodily location, gesture, or immediate physical sensation rather than through a fully specified verbal description (e.g., \textit{``It hurts here''}).
Finally, \textbf{Certainty} captures the patient's confidence in the information provided as \textit{certain}, \textit{approximate}, or \textit{doesn't know}. Certain responses contain no hedging (e.g., \textit{``It started three days ago''}); approximate responses contain hedging while still providing an answer (e.g., \textit{``Maybe three or four days ago''}); and doesn't know indicates an explicit inability to recall.

\textbf{Physical Examination (Conditional)}: When conversational turns related to physical examination are identified, we apply an additional set of L3 markers. Physical examination differs from other consultation phases because the interaction involves translating physical actions and observations into verbal conversation.
\textbf{Deictic Reference Density} measures the frequency of references to physical locations or actions (e.g.: \textit{``Does it hurt when I press here?''} ). 
\textbf{Narration Density} measures the proportion of examination turns in which the doctor explicitly verbalizes what they are doing or are about to do (e.g., \textit{``I'm going to press your abdomen now''}).
\textbf{Findings Disclosure} measures the proportion of examination turns in which the doctor explicitly states what they observed (e.g., \textit{``Breath sounds are reduced on the right''}). 
\textbf{Compliance} is assigned to patient examination turns as a binary label: \textit{compliant} when the patient follows the instruction without additional contribution (e.g., Doctor: \textit{``Take a deep breath''} $\rightarrow$ Patient: \textit{``Okay''}) and \textit{proactive} when the patient follows the instruction while additionally volunteering information (e.g. \textit{``Yes, and it also hurts a bit to the right''}).
\section{Experimental Setup} 
\begin{table*}[!t]
    \centering
    \setlength{\tabcolsep}{2pt}
    \caption{Datasets used in the evaluation. \textit{Real} = recorded consultations; \textit{Simulated} = role-played or annotator-written; \textit{Synthetic} = LLM-generated. en = English, hi = Hindi, te = Telugu. \textit{Notation} denotes the shorthand to refer to each dataset throughout the study.}
    \label{tab:datasets}
    \small
    % \renewcommand{\arraystretch}{1.3}
    % \begin{tabular}{@{} p{0.12\linewidth}
    %                     p{0.12\linewidth}
    %                     p{0.08\linewidth}
    %                     p{0.08\linewidth}
    %                     p{0.50\linewidth} @{}}
    \begin{tabular}{lcccp{0.5\linewidth}}
    \toprule
    \textbf{Dataset} & \textbf{Notation} & \textbf{Lang.} & \textbf{Type} & \textbf{Description} \\
    \midrule

    MTS-Dialog~\cite{ben-abacha-etal-2023-empirical}
      & Sim-US-1
      & en & Simulated
      & 1.7k doctor-patient dialogues manually written by healthcare professionals from de-identified real clinical notes \\
    
    ACI-Bench~\cite{yim2023acibench}
      & Sim-US-2
      & en & Simulated
      & 207 doctor-patient encounters constructed by domain experts through role-play each paired with a reference clinical note \\
    
    NoteChat~\cite{wang-etal-2024-notechat}
      & Syn-US-1
      & en & Synthetic
      & GPT-3.5 multi-agent pipeline grounded in PMC case reports \\
    
    Clinical Note Generation ~\cite{ekacare2025clinicalnote}
      & Sim-Indian
      & hi & Simulated
      & 156 transcribed consultations between EkaCare's clinicians and internal staff who role-played as patients, each annotated with structured ground-truth notes. \\
    
    Privtel
      & Real-Indian
      & te & Real
      & Proprietary corpus of real outpatient consultations \\
    
    \midrule
    \multicolumn{5}{@{}l}{{Synthetic Generation Dataset,
    40 samples per cell per language}} \\
    \midrule
    
    \multirow{3}{*}{Free-form, ungrounded}
      & Syn-US-2
      & en & \multirow{3}{*}{Synthetic}
      & \multirow{3}{*}{Single prompt; model generates an outpatient conversation} \\

      & Syn-Indian-1
      & hi & 
      & \\

      &Syn-Indian-2
      & te & 
      & \\
    
    \multirow{3}{*}{Free-form, MTS-grounded}
      & Syn-US-2
      & en & \multirow{3}{*}{Synthetic}
      & \multirow{3}{*}{Single prompt; model seeded with an MTSamples clinical note} \\
      
      & Syn-Indian-1
      & hi & 
      & \\

      &Syn-Indian-2
      & te & 
      & \\
      
    \multirow{3}{*}{Agentic, ungrounded}
      & Syn-US-2
      & en & \multirow{3}{*}{Synthetic}
      & \multirow{3}{*}{Doctor- and patient-agent; no case grounding; $\geq$20 turns} \\
      
      & Syn-Indian-1
      & hi & 
      & \\

      &Syn-Indian-2
      & te & 
      & \\
      
    \multirow{3}{*}{Agentic, MTS-grounded}
      & Syn-US-2
      & en & \multirow{3}{*}{Synthetic}
      & \multirow{3}{=}{Doctor- and patient-agent; MTSamples notes used as shared grounding; $\geq$20 turns} \\

      & Syn-Indian-1
      & hi & 
      & \\

      &Syn-Indian-2
      & te & 
      & \\
    
    \bottomrule
    \end{tabular}
\end{table*}

\subsection{Datasets}
To answer our research questions, we identify datasets (summarized in Table~\ref{tab:datasets}) that provide contrasts along the two dimensions: the cultural setting of the consultation (India vs. US) and how the conversation is generated. 
% whether from transcripts of naturally occurring interactions (real), or written by clinical experts from patient history (simulated), or LLM-generated dialogue (synthetic). 
% We therefore bring together existing datasets that provide complementary views of these dimensions, .
We consider three sources of clinical dialogue: \textit{real} consultations, consisting of real-world doctor-patient interactions; \textit{simulated} consultations, constructed either as scripted transcripts written by clinical experts or as enacted role-plays between clinicians and trained actors; 
% produced through clinical expert written scribes or role-play rather than recorded clinical encounters; 
and \textit{synthetic} consultations, generated using language models, at times grounded in a real patient case. 

For the \textit{real} Indian setting, we use a private corpus of Telugu outpatient consultations collected by a healthcare provider, which we refer to as PrivTel. We could not identify an openly available US corpus that provides a comparable real-world counterpart. For cross-cultural contrast, we treat the simulated US datasets as proxies for Western clinical interaction and compare them with Indian simulated counterparts. 

% , providing the real Indian reference point for our framework.
For the \textit{simulated} setting, we use MTS-Dialog \cite{ben-abacha-etal-2023-empirical} and ACI-Bench \cite{yim2023acibench} as US consultation counterparts, and Clinical Note Generation \cite{ekacare2025clinicalnote} as an Indian counterpart (Table \ref{tab:datasets}).
% Skipping this, added to the table
MTS-Dialog~\cite{ben-abacha-etal-2023-empirical} comprises $\sim$1.7k doctor-patient dialogues manually written by healthcare professionals from de-identified clinical notes, while ACI-Bench~\cite{yim2023acibench} comprises 207 doctor-patient encounters constructed by domain experts through role-play, each paired with a reference clinical note. Clinical Note Generation~\cite{ekacare2025clinicalnote} comprises 156 transcribed consultations between EkaCare clinicians and internal staff who role-played as patients, in English, Hindi, and Marathi, each annotated with structured, entity-level ground-truth notes.
For the \textit{synthetic} setting, we use NoteChat \cite{wang-etal-2024-notechat} as a US reference;
% due to its popularity as a fully synthetic dataset; 
to the best of our knowledge, no synthetic reference dataset is available from India.
For each reference dataset, we randomly select 150 conversations for analysis.
% We used a similar annotation scale to both MTS-Dialog~\cite{ben-abacha-etal-2023-empirical} (100) and ACI-Bench~\cite{yim2023acibench} (163). 
% This scale follows evaluation practice in comparable annotation studies~\cite{ben-abacha-etal-2023-empirical,yim2023acibench} 
% to observe comparable qualitative differences that the framework targets. 
All the datasets consist of general outpatient consultations rather than specialist or emergency encounters, providing a comparable clinical setting, and are stratified by patient gender.

We additionally construct a synthetic corpus across two generation modes and two clinical-grounding conditions, reflecting commonly used approaches to generating synthetic clinical consultations~\cite{wang-etal-2024-notechat, das2024synthetic}. The generation modes and prompt instructions are intentionally minimal and open-ended, allowing us to examine what the model considers culturally appropriate without explicit rules or fine-grained tuning. Complete prompts are provided in the Appendix. \textit{Free-form} generation uses a single agent to produce the consultation, while \textit{agentic} generation uses separate doctor and patient agents that interact for at least 20 turns. Within each mode, conversations are either \textit{ungrounded} or \textit{MTS-grounded}, with the latter conditioned on a randomly sampled subset of MTSamples\footnote{\url{https://mtsamples.com}: Public repository of de-identified clinical transcription notes.}. We generate 40 conversations per condition, resulting in 160 synthetic conversations for each of English, Hindi, and Telugu.
% which matches the reference corpus's dataset size. 
For each language, the model is instructed to generate a consultation situated in the appropriate clinical and cultural setting, with culturally appropriate doctor and patient behavior. For English, this corresponds to a US outpatient setting, while for Hindi and Telugu, the model is instructed to generate an Indian outpatient consultation. The Indian language prompts also require that conversations be generated in the respective native scripts. 

% Add prompts and example generations in appendix. Maybe I can also add iterations (plain vs giving cultural roles)

\subsection{Miscellaneous Details}
\textbf{Models.}
We use Gemini 3.7 Flash \cite{google2026gemini37flash} for dialogue generation of the synthetic corpus described above, chosen for its multilingual instruction-following capabilities and support for multi-turn generation, which is required for the agentic setting. For evaluation, we use GPT-5.1 \cite{openai2025gpt51} for LLM-based annotation,  with a temperature of 0. All prompts, including their few-shot examples, are provided in the supplementary material.
% and score the resulting conversations. 

\noindent \textbf{Human Evaluation.}
We conduct a human evaluation to assess whether the proposed interactional cultural markers (Table~\ref{tab:framework}) can be reliably identified from clinical conversations and to validate the corresponding LLM-based annotations. We sample 10 conversations from each language, equally split across the available settings (e.g.: English has only simulated and synthetic datasets, so we sample five conversations from each setting). Two human annotators independently apply the same layer-by-layer annotation scheme used by the evaluation model. We first measure inter-annotator agreement to assess consistency between the two human annotators. Cohen's $\kappa$ values of $0.71$ (en), $0.68$ (hi), and $0.72$ (te) indicate substantial agreement between the annotators. We then compare the human annotations with the LLM-based annotations to assess whether the LLM capture the same interactional patterns. 
Cohen's $\kappa$ values between the averaged human annotations and LLM-based annotations are $0.846$ (en), $0.813$ (hi), and $0.804$ (te), respectively. These substantial levels of agreement support the use of LLM-based annotations for the subsequent analysis.
% Averaged human annotations agree with LLM-based annotations result in agreement on 84.6\% (en), 81.3\% (hi), and 80.4\% (te) patterns, respectively. 
% assess whether the LLM evaluation captures patterns identified by human annotators.

\noindent \textbf{Usage of AI.}
In this work, we use LLMs for synthetic data generation, as described above; for evaluation through LLM-based annotation; and as coding assistants to develop the evaluation code. All outputs generated by these models were verified by two of the authors.
\section{Findings}
We structure our findings around the three research questions, progressively extending the analysis from observed interactional differences to their reproduction in synthetic dialogue. We first examine interactional differences between Indian and US consultations, using simulated US clinical coversations as a contrast, because we could not identify a comparable real-world US corpus. The US datasets serve as a reference point for situating the interactional characteristics of Indian conversations; we therefore ground observable properties in literature but do not prescribe how US clinical dialogue should be generated or modeled.
We then examine whether the identified characteristics transfer to synthetic clinical dialogue. Finally, we analyze how this fidelity varies across languages, generation settings, and patient characteristics, within the synthetic data.

\begin{figure*}[t]
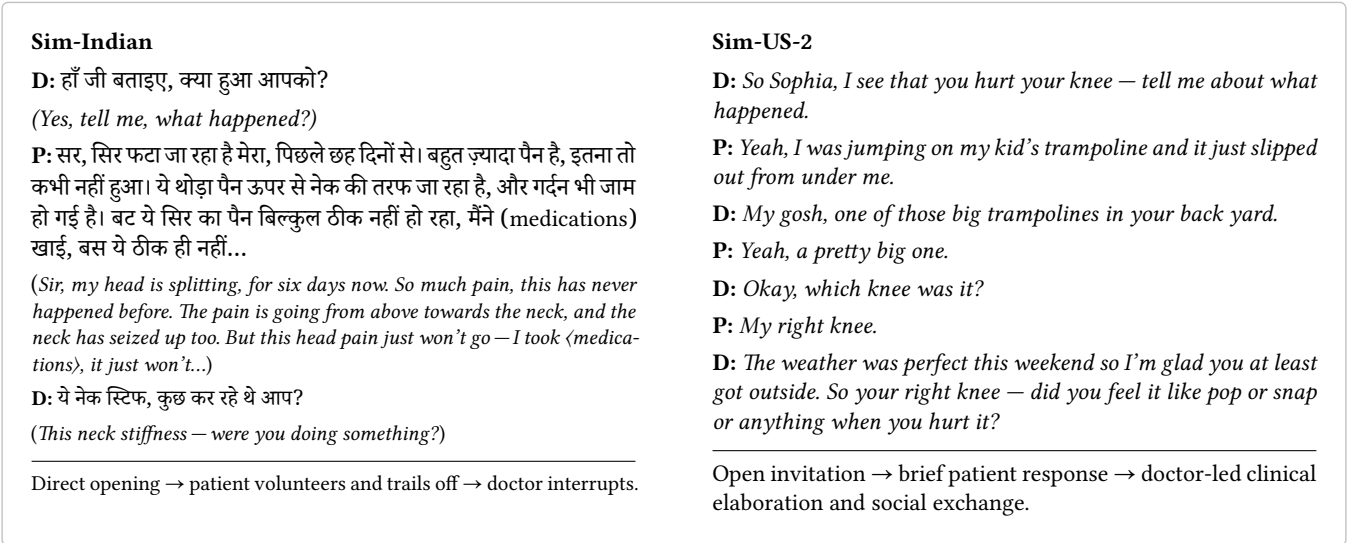

\centering
\begin{tcolorbox}[
    colback=white,
    colframe=black!25,
    boxrule=0.5pt,
    arc=2pt,
    width=\textwidth,
    left=8pt,
    right=8pt,
    top=8pt,
    bottom=8pt
]

\begin{minipage}[t]{0.47\textwidth}
\textbf{Sim-Indian}

\vspace{4pt}

\textbf{D:} {\devanagarifont हाँ जी बताइए, क्या हुआ आपको?}
\vspace{3pt}

\textit{(Yes, tell me, what happened?)}

\vspace{3pt}

\textbf{P:} {\devanagarifont सर, सिर फटा जा रहा है मेरा, पिछले छह दिनों से। बहुत ज़्यादा पैन है, इतना तो कभी नहीं हुआ। ये थोड़ा पैन ऊपर से नेक की तरफ जा रहा है, और गर्दन भी जाम हो गई है। बट ये सिर का पैन बिल्कुल ठीक नहीं हो रहा, मैंने ({\normalfont medications})  खाई, बस ये ठीक ही नहीं{\dots}}

\vspace{3pt}

(\small\textit{Sir, my head is splitting, for six days now. So much pain, this has never happened before. The pain is going from above towards the neck, and the neck has seized up too. But this head pain just won't go\,---\,I took \textlangle{}medications\textrangle{}, it just won't{\dots}})

\vspace{3pt}

\textbf{D:} {\devanagarifont ये नेक स्टिफ, कुछ कर रहे थे आप?}
\vspace{3pt}

(\textit{This neck stiffness\,---\,were you doing something?})

\vspace{6pt}
\hrule
\vspace{5pt}

Direct opening $\rightarrow$ patient volunteers and trails off $\rightarrow$ doctor interrupts.

\end{minipage}
\hfill
\begin{minipage}[t]{0.47\textwidth}
\textbf{Sim-US-2}

\vspace{4pt}

\textbf{D:} \textit{So Sophia, I see that you hurt your knee --- tell me about what happened.}

\vspace{3pt}

\textbf{P:} \textit{Yeah, I was jumping on my kid's trampoline and it just slipped out from under me.}

\vspace{3pt}

\textbf{D:} \textit{My gosh, one of those big trampolines in your back yard.}

\vspace{3pt}

\textbf{P:} \textit{Yeah, a pretty big one.}

\vspace{3pt}

\textbf{D:} \textit{Okay, which knee was it?}

\vspace{3pt}

\textbf{P:} \textit{My right knee.}

\vspace{3pt}

\textbf{D:} \textit{The weather was perfect this weekend so I'm glad you at least got outside. So your right knee --- did you feel it like pop or snap or anything when you hurt it?}

\vspace{6pt}
\hrule
\vspace{5pt}

Open invitation $\rightarrow$ brief patient response $\rightarrow$ doctor-led clinical elaboration and social exchange. 

\end{minipage}
\end{tcolorbox}

\caption{Contrasting interactional organisation in Indian and US clinical conversations.}
\label{fig:indian-english-consultation}
\end{figure*}

\subsection{RQ1: What interactional characteristics distinguish Indian consultations from US clinical contexts?}
Across the two Indian datasets (Real-Indian and Sim-Indian), despite differences in language and data collection settings, we observe a recurring pattern of clinical interaction that differs from the simulated US conversations.
% The consistent trend of these patterns across datasets suggests that they reflect broader interactional tendencies rather than properties of any single dataset.
Figure \ref{fig:indian-english-consultation} illustrates this contrast. In the Indian example, a short, direct opening elicits an extended patient account containing local jargon, which the doctor then interrupts to redirect toward the immediate clinical topic. In the US example, the doctor opens with a more discussive prompt, but the patient response is brief; the doctor subsequently contributes through social exchange and clinical elaboration before continuing with focused questioning. The aggregate markers in Table \ref{tab:dataset-characteristics} show that these patterns extend beyond isolated examples.

\begin{table*}[ht]
\centering
\caption{Key interactional cultural markers averaged for each dataset. \textbf{D:P ratio}: doctor-to-patient word ratio; values below 1.0 indicate patient-dominant speech. \textbf{HT:Tx ratio}: history-taking turns per treatment turn; \textbf{Directiveness}: proportion of all doctor turns that are directive. \textbf{Symptom Description}: proportion of patient turns containing medical or technical vocabulary. \textbf{Response-Agreement}: proportion of patient turns that agree without asking a follow-up question. \textbf{Trailing Responses}: proportion of patient responses that extend beyond the immediate question or topic.}
\small
\setlength{\tabcolsep}{5pt}
% \resizebox{\textwidth}{!}{%
\begin{tabular}{lccccccc}
\toprule
\textbf{Dataset} &
\makecell{\textbf{D:P}\\\textbf{ratio}} &
\makecell{\textbf{HT:Tx}\\\textbf{ratio}} &
\makecell{\textbf{Directiveness}\\\textbf{[D]}} &
\makecell{\textbf{Symptom}\\\textbf{Description [P]}} &
\makecell{\textbf{Response-}\\\textbf{Agreement [P]}} &
\makecell{\textbf{Trailing}\\\textbf{Responses [P]}} &
\makecell{\textbf{Interruption}\\\textbf{Rate [D]}} \\
\midrule
Real-Indian              & 0.79 & 6.5 & 58\% & 13\% & 72\% & 59\% & 23\% \\
Sim-Indian               & 0.89 & 4.2 & 62\% & 16\% & 64\% & 48\% & 12\% \\

\midrule

Sim-US-1            & 2.06 & 1.5 & 20\% & 28\% & 18\% & 24\% & $\sim$0\% \\
Sim-US-2            & 1.95 & 1.4 & 18\% & 24\% & 21\% & 22\% & $\sim$0\% \\

\midrule

Syn-US-1            & 2.87 & 1.2 & 26\% & 32\% & 26\% & 27\% & $\sim$0\% \\
Syn-US-2            & 2.67 & 1.4 & 24\% & 36\% & 20\% & 16\% & $\sim$0\% \\
Syn-Indian-1             & 2.47 & 1.1 & 28\% & 28\% & 17\% & 14\% & $\sim$0\% \\
Syn-Indian-2             & 2.59 & 0.9 & 22\% & 34\% & 15\% & 8\% & $\sim$0\% \\
\bottomrule
\end{tabular}%
% }
\label{tab:dataset-characteristics}
\end{table*}
\textbf{Conversational dominance.} The \textit{D:P ratio}, computed from Word Count (L1), shows that the simulated US conversations are doctor-dominant (average ratio of 2.01), whereas both Indian datasets are patient-dominant (average ratio of 0.84). The \textit{Directiveness} measure, computed from Tone (L3), provides a complementary perspective: Indian doctors are considerably more directive (58\% and 62\% vs. 20\% and 18\% in US conversations). Indian doctors typically use short, bounded questions to guide patients through specific aspects of their history, even as patients contribute more of the overall speech. The lower \textit{HT:Tx ratio} further indicates that Indian consultations devote substantially more conversational space to history-taking before relative to treatment. 

\textbf{Participation and interaction.} The \textit{Response-Agreement} measure, computed from Response Reaction (L3), and \textit{Trailing Responses}, computed from Response Style (L3), point to a distinct pattern of patient participation in the Indian consultations. Patients frequently provide information beyond what is explicitly requested, but are less likely to ask questions of their own and show substantially higher agreement rate (72\% and 64\% vs. 18\% and 21\%). These extended responses are often accompanied by interruptions (\textit{Interruption Rate} (L2)), with doctors redirecting patients back to the immediate topic. The simulated US consultations, by contrast, show less trailing and interruption, but more explicit negotiation through questions, follow-up, and responses to doctors' contributions. These observations are consistent with prior literature characterizing Indian consultations as more directive and US consultations as involving greater discussion and patient participation~\cite{mehradoctorinfluence,riasroter}.

\textbf{Linguistic register.} Indian patients more often describe illness through colloquial and locally familiar expressions, as illustrated in Figure~\ref{fig:indian-english-consultation} ({\devanagarifont सिर फटा जा रहा है}, `head is splitting' is a local slang for headache). Code-switching often appears around clinical concepts (e.g., {\devanagarifont बहुत ज़्यादा पेन {\normalfont(pain)} है}, `there's a lot of pain'), with formal medical framing introduced progressively throughout the interaction. The clinical history in Indian consultations therefore often begins in an everyday linguistic register, with patients' own phrasing becoming part of how the history is constructed. US consultations contain substantially more formal clinical terminology in patient speech, as reflected in \textit{Symptom Description} (L3) (28\% and 24\% vs. 13\% and 16\% in Indian setting).

\subsection{RQ2: How well does synthetic clinical dialogue capture the cultural characteristics?}

\begin{figure*}[t]
\centering
\begin{tcolorbox}[
    colback=white, colframe=black!25, boxrule=0.5pt,
    arc=2pt, width=\textwidth, left=6pt, right=6pt, top=6pt, bottom=6pt
]

\begin{minipage}[t]{0.47\textwidth}
\textbf{Sim-Indian}

\vspace{4pt}
\textbf{D:} {\devanagarifont कितना सीवियर हो रहा है सर दर्द?}

\vspace{2pt}\small\textit{(How severe is the headache?)}

\vspace{3pt}
\textbf{P:} {\devanagarifont आज तो बहुत ज़्यादा हो रहा है। बट इन जनरल मुझे ऐसे कभी ना कभी हो जाता है। बट हाँ, सुबह से अभी तक बंद नहीं हुआ है।}

\vspace{2pt}\small\textit{(Today it's a lot. But in general it happens to me sometimes. But yes, since morning it hasn't stopped.)}

\vspace{3pt}
\textbf{D:} {\devanagarifont इसके साथ बुखार या उल्टी कुछ हो रहा है?}

\vspace{2pt}\small\textit{(Any fever or vomiting with it?)}

\vspace{3pt}
\textbf{P:} {\devanagarifont नहीं नहीं। थोड़ा वॉमिट टाइप फ़ील हो रहा है बट कुछ हुआ नहीं है।}

\vspace{2pt}\small\textit{(No no. Feeling a little vomit-type but nothing has happened.)}

\vspace{3pt}
\textbf{D:} {\devanagarifont ठीक है, दवाई लिखता हूँ। पाँच दिन लो फिर आना। स्ट्रेस रिलेटेड हो सकता है या माइग्रेन।}

\vspace{2pt}\small\textit{(Okay, I'll prescribe medicine. Take it for five days then come back. Could be stress-related or migraine.)}

\vspace{6pt}\hrule\vspace{4pt}

\small\textbf{Pattern:} Short, terse questions in one turn; patient elaborates beyond the question; no reassurance.

\end{minipage}
\hfill
\begin{minipage}[t]{0.47\textwidth}
\textbf{Syn-Indian-1}

\vspace{4pt}
\textbf{D:} {\devanagarifont बताइए — दर्द कितने समय से है? नींद पर असर पड़ रहा है? कोई पुरानी बीमारी है? पानी पर्याप्त पी रहे हैं?}

\vspace{2pt}\small\textit{(Tell me --- how long has the pain occured? Sleep affected? Pre-existing conditions? Drinking enough water?)}

\vspace{3pt}
\textbf{P:} {\devanagarifont जी डॉक्टर साहब, दो हफ्ते से है। हाँ सर, नींद बहुत अफेक्ट हुई है। नहीं डॉक्टर साहब, कोई पुरानी बीमारी नहीं है। पानी थोड़ा कम पीता हूँ सर।}

\vspace{2pt}\small\textit{(Yes doctor sahab, two weeks. Yes sir, sleep very affected. No doctor sahab, no conditions. I drink less water sir.)}

\vspace{3pt}
\textbf{D:} {\devanagarifont देखिए, आपके सिम्टम्स के आधार पर यह माइग्रेन हो सकता है — लेकिन चिंता मत कीजिए, यह एक मैनेजेबल कंडीशन है और ट्रीटमेंट से पूरी तरह ठीक हो सकता है।}

\vspace{2pt}\small\textit{(See, based on your symptoms this could be migraine --- but don't worry, it is a manageable condition and can be fully treated.)}

\vspace{3pt}
\textbf{P:} {\devanagarifont डॉक्टर साहब बहुत-बहुत धन्यवाद। क्या यह सीरियस तो नहीं है सर?}

\vspace{2pt}\small\textit{(Thank you so much doctor sahab. This isn't serious, is it sir?)}

\vspace{6pt}\hrule\vspace{4pt}
\small\textbf{Pattern:} Structured questioning and symptom listing, patient addresses doctor in every turn; empathetic framing.

\end{minipage}

\end{tcolorbox}
\caption{Simulated vs.\ synthetic Hindi consultations highlighting signatures present in generations.}
\label{fig:hindi-contrast}
\end{figure*}

\textbf{Real-to-simulated-to-synthetic progression.} Comparing the three Indian settings reveals a gradual shift in the representation of clinical interaction. The Real-Indian (te) and Sim-Indian (hi) conversations remain broadly similar in their overall profile, but the simulated conversations smooth over some of the interactional variation present in the real conversations. For example, patients in the real conversations more often extend their responses beyond the immediate question (59\% vs. 48\% in Trailing Responses (L3)) and respond with agreement without further questioning (72\% vs. 64\% in Response-Agreement (L3)), while interruptions occur almost twice as often (23\% vs. 12\% in Interruption Rate (L2)). These differences suggest that simulation captures the general pattern of patient participation and doctor direction, but captures less effectively the bounded and variable exchanges through which these patterns emerge in real consultations. The shift becomes more pronounced in Syn-Indian-1 (hi), where interruptions are effectively absent (0\%), patient responses become more certain (74.2\%), and doctors become substantially more verbose (28.5\%) and empathetic (38.2\%).
% , and structurally elaborate in their questioning.

\textbf{Alignment with US interactional patterns.} Across the four synthetic datasets, the consultation characteristics more closely resemble the US data than the Indian data, irrespective of whether the conversations are prompted as Indian or US consultations (Table \ref{tab:dataset-characteristics}). Synthetic conversations are substantially more doctor-dominant, devote less conversational space to history-taking, and exhibit substantially less directive questioning and interruption than observed in the real Indian consultations. These patterns suggest that synthetic clinical dialogue generation defaults to interactional patterns closer to those observed in the US clinical data, despite explicit prompting to reproduce Indian language and cultural context. One possible explanation is that LLMs are disproportionately exposed to English-language and US-centric clinical communication during training.

\begin{table*}[ht]
\centering
\caption{Comparison of markers between synthetic consultations and their real or simulated counterparts.
\textbf{Vocative Frequency}: proportion of patient turns containing a direct address term (sir, doctor, {\devanagarifont सर}, ayya, sahab).
\textbf{Register}: proportion of doctor turns using clinical or clinical-vernacular terminology.
\textbf{Empathetic}: proportion of doctor turns containing explicit empathetic or rapport-building language ("Don't worry", "I understand").
\textbf{Verbosity}: proportion of doctor turns rated as over-explanatory.
\textbf{Question-Structure}: proportion of doctor turns containing $\geq$2 topically unrelated questions.
\textbf{Certainty}: proportion of confident patient responses with no uncertain or doubtful signatures.}
\small
% \resizebox{\textwidth}{!}{%
\begin{tabular}{lcccccc}
\toprule
\textbf{Dataset} & \textbf{\begin{tabular}[c]{@{}c@{}}Vocative \\ Frequency {[}P{]}\end{tabular}} & \textbf{\begin{tabular}[c]{@{}c@{}}Clinical \\ Register {[}D{]}\end{tabular}} & \textbf{\begin{tabular}[c]{@{}c@{}}Tone-\\ Empathetic {[}D{]}\end{tabular}} & \textbf{\begin{tabular}[c]{@{}c@{}}Verbosity\\ {[}D{]}\end{tabular}} & \textbf{\begin{tabular}[c]{@{}c@{}}Question\\ Structure {[}D{]}\end{tabular}} & \textbf{\begin{tabular}[c]{@{}c@{}}Certainty\\ {[}P{]}\end{tabular}} \\ 
% \textbf{Dataset} &
% \textbf{Vocative Frequency [P]} &
% \textbf{Clinical Register [D]} &
% \textbf{Tone-Empathetic [D]} &
% \textbf{Verbosity [D]} &
% \textbf{Question Structure [D]} &
% \textbf{Certainty [P]} \\
\midrule
Sim-Indian       & 15.6\% & 38.2\% &  4.3\% &  9.1\% &  1.2\% & 38.4\% \\
Syn-Indian-1     & 33.3\% & 55.3\% & 38.2\% & 28.5\% & 19.1\% & 74.2\% \\
\midrule
Sim-US-1    &  2.2\% & 62.4\% & 18.3\% & 16.2\% &  1.7\% & 58.1\% \\
Syn-US-1    &  7.2\% & 74.3\% & 25.2\% & 35.4\% &  4.0\% & 82.3\% \\
Syn-US-2    & 12.3\% & 68.2\% & 32.4\% & 38.1\% &  8.6\% & 85.2\% \\
\midrule
Real-Indian      & 16.7\% & 35.2\% &  3.3\% &  7.2\% &  1.8\% & 32.4\% \\
Syn-Indian-2     & 36.4\% & 52.4\% & 41.2\% & 32.3\% & 10.9\% & 79.3\% \\
\bottomrule
\end{tabular}%
% }
\label{tab:pragmatic-features}
\end{table*}

\textbf{Synthetic signatures.} Beyond these broader interactional differences, synthetic data exhibit recurring patterns across languages. Figure \ref{fig:hindi-contrast} shows that Syn-Indian-1 (hi) uses repeated address terms, reassurance, empathy, and elaborated questioning, in contrast to the terse and task-focused Sim-Indian exchange. Table \ref{tab:pragmatic-features} shows that these patterns generalizes across synthetic consultations. One prominent signature is the increased use of direct address terms by patients (Vocative Frequency (L1)), with vocatives such as `sir' or `doctor' substantially more frequent (33.3-36.4\% vs. 15.6-16.7\% for Indian data; 7.2-12.3\% vs. 2.2\% for US data). Doctors also use a more formal Clinical Register (L3) and are substantially more Verbose (L3), providing more elaborate explanations than those observed in the real conversations. Synthetic doctors also use considerably more explicit empathy and rapport-building language, with Empathetic (L3) language rising from 3.3-4.3\% to 38.2-41.2\% in Indian consultations, and from 18.3\% to 25.2-32.4\% in US consultations. Questions are also more structurally elaborate (Question Structure (L3)), with synthetic doctors more often combining multiple clinical questions within a single turn (Figure~\ref{fig:hindi-contrast}, Syn-Indian-1 examples). On the patient side, higher Certainty (L3) of responses suggests another characteristic of generated dialogue: patients tend to provide clean and confident answers rather than the hesitation, uncertainty, or qualification observed in clinical histories.

\begin{figure*}[t]
\centering
\begin{tcolorbox}[
    colback=white, colframe=black!25, boxrule=0.5pt,
    arc=2pt, width=\textwidth,
    left=6pt, right=6pt, top=6pt, bottom=6pt
]

\small
\renewcommand{\arraystretch}{1.15}

\begin{tabularx}{\textwidth}{@{}X X X@{}}

\textbf{Syn-US-1 and Syn-US-2} &
\textbf{Syn-Indian-1} &
\textbf{Syn-Indian-2}
\\
\midrule

\multicolumn{3}{@{}l}{\textit{\textbf{(i) Unprompted doctor reassurance}}}
\\[4pt]

\textbf{D:} \textit{Based on your symptoms, this appears to be a urinary tract infection. I'll prescribe antibiotics.}
&
\textbf{D:} {\devanagarifont आपकी फीलिंग्स बिल्कुल वैलिड हैं — लेकिन चिंता मत कीजिए, हम मिलकर इसे मैनेज करेंगे।}
\textit{(Your feelings are completely valid — don't worry, we'll manage this together.)}
&
\textbf{D:} {\telugufont భయపడకండి సర్, ఇది చాలా కామన్ కండిషన్. మీరు సరైన సమయానికి వచ్చారు.}
\textit{(Don't be afraid sir, this is very common. You came at the right time.)}
\\[6pt]

\midrule

\multicolumn{3}{@{}l}{\textit{\textbf{(ii) Patient over-compliance}}}
\\[4pt]

\textbf{P:} \textit{Yes, that's correct.}
&
\textbf{P:} {\devanagarifont हाँ सर, आपने बिल्कुल सही कहा। मैं ज़रूर फॉलो करूँगा डॉक्टर साहब।}
\textit{(Yes sir, absolutely right. I will definitely follow it doctor sahab.)}
&
\textbf{P:} {\telugufont సరే డాక్టర్ సాహబ్, మీరు చెప్పినట్టే చేస్తాను. మీకు చాలా కృతజ్ఞుడిని సర్.}
\textit{(Okay doctor sahab, I will do as you say. I am very grateful to you sir.)}
\\[6pt]

\midrule

\multicolumn{3}{@{}l}{\textit{\textbf{(iii) Register leakage}}}
\\[4pt]

\textbf{P:} I've been experiencing \underline{nausea} and \underline{fatigue} for the past few days.
\textit{[mildly formal, within lay register]}
&
\textbf{P:} {\devanagarifont नहीं, मुझे \underline{श्वसन} में दिक्कत नहीं है, लेकिन \underline{हृदय स्पंदन} तीव्र हो जाता है।}
\textit{(No respiratory difficulty, but cardiac rhythm becomes rapid.)}
\newline
\textit{[lay: {\devanagarifont सांस ठीक है, पर घबराहट होने लगती है।} --- ``Breath fine, but fluttery feeling.'']}
&
\textbf{P:} {\telugufont సర్, నాకు \underline{హైపర్టెన్షన్} వల్ల \underline{కార్డియాక్ అరెస్ట్} అవుతుందా?}
\textit{(Sir, can hypertension cause cardiac arrest?)}
\newline
\textit{[lay: {\telugufont బీపీ ఎక్కువైతే గుండెకి ఏమైనా అయిపోతుందా?} --- ``If BP goes up, will something happen to my heart?'']}
\\

% \bottomrule

\end{tabularx}

\end{tcolorbox}

\caption{Recurrent Indian-prominent synthetic artefacts across language groups. \textit{Lay} refers to what a typical layman response in the same context would look like without the use of heavy clinical vocabulary.}
\label{fig:synthetic-artefacts}
\end{figure*}

\textbf{Indian-specific synthetic signatures.} In addition to these general synthetic patterns, several markers are disproportionately prominent in the synthetic Indian consultations. Examples in Figure \ref{fig:synthetic-artefacts} illustrates these patterns. Syn-Indian-1 (hi) and Syn-Indian-2 (te) repeatedly introduce reassurance without an apparent conversational trigger, while patients frequently respond with gratitude and deferential address terms. Doctors also occasionally address patients using kinship terms such as {\devanagarifont भाई} (\textit{bhai}, brother), {\devanagarifont बहन} (\textit{behen}, sister), or {\devanagarifont बेटा} (\textit{beta}, son), including cases where these terms do not correspond to the patient's gender or age.
Synthetic Indian conversations frequently express clinical concepts using formal medical vocabulary, in sharp contrast to the real-world Indian conversations.
% rather than the more varied and conversational descriptions observed in the underlying conversations. 
Another pattern is the use of highly Sanskritised expressions (underlined text in Figure \ref{fig:synthetic-artefacts}) alongside English medical terminology, producing forms of code-mixing that differ from the mixture of colloquial and clinical terminology in the real-world conversations. 

% Synthetic patients also show more pronounced register leakage, using highly technical or Sanskritised expressions where everyday local speech would be more natural. Although the reason for these associations is unclear, the pattern suggests that the generation process may be linking Indian clinical settings with heightened reassurance, gratitude, and culturally marked forms of address, producing these features more strongly than they occur in the underlying consultations.

% \input{figures/physicalexamination}

\begin{figure}[t]
\centering
\includegraphics[
    width= 0.45\textwidth
]{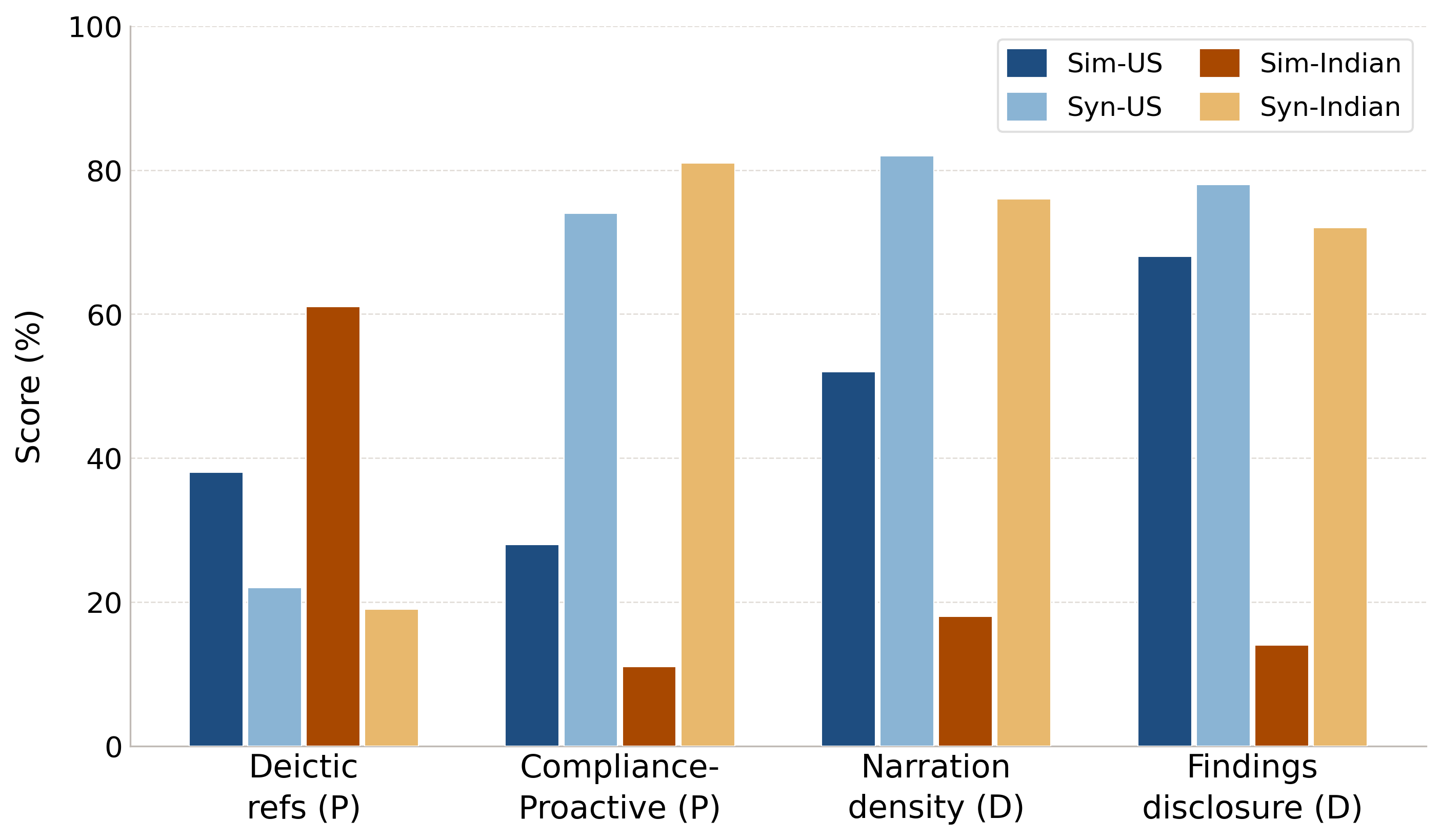}
\caption{Quantitative comparison of physical-examination dialogue patterns averaged across simulated, and synthetic consultations. \textbf{Deictic References}: context-dependent expressions like '' here,'' or ``on this side'' \textbf{Narration Density} verbal description of examination actions; \textbf{Compliance- Proactivity}: Percentage of turns where patient contributes more than what the doctor asks \textbf{Findings Disclosure}: Verbalization of examination-related observations.}
\label{fig:phys_exam_metrics}
\end{figure}

\begin{figure*}[t]
\centering
\begin{tcolorbox}[
    colback=white,
    colframe=black!25,
    boxrule=0.5pt,
    arc=2pt,
    width=\textwidth,
    left=6pt,
    right=6pt,
    top=6pt,
    bottom=6pt
]

\begin{minipage}[t]{0.32\textwidth}
    \footnotesize
    \raggedright

    \textbf{Sim-US-1}

    \vspace{2pt}

    \textbf{D:} \textit{Okay. Back up here in the stretcher a bit more. Push your leg out against my hand --- does that hurt?}

    \textbf{D:} \textit{And if you pull back?}

    \textbf{D:} \textit{Does it hurt more on the inside or more on the outside here?}

    \textbf{P:} \textit{The right knee here --- on the outside.}

    \textbf{D:} \textit{And here, any numbness or tingling?}

    \textbf{P:} \textit{No.}

\end{minipage}
\hfill
\begin{minipage}[t]{0.32\textwidth}
    \footnotesize
    \raggedright

    \textbf{Syn-US-1}

    \vspace{2pt}

    \textbf{D:} \textit{Okay, I'd like to go ahead and do a physical examination. Lie back for me.}

    \textbf{D:} \textit{I'm going to press on your abdomen. Any pain there?}

    \textbf{P:} \textit{Yeah --- and actually a bit to the right too, not just there.}

    \textbf{D:} \textit{There is tenderness on palpation of the right lower quadrant. Any pain when I let go?}

    \textbf{P:} \textit{Yeah, that's actually sharper --- when you release.}

\end{minipage}
\hfill
\begin{minipage}[t]{0.32\textwidth}
    \footnotesize
    \raggedright

    \textbf{Syn-Indian-1}

    \vspace{2pt}

    \textbf{D:} {\devanagarifont \textit{ठीक है, अब चेस्ट एग्ज़ामिन करूँगा। गहरी साँस लीजिए।}}

    \vspace{1pt}

    {\small
    (\textit{\textbf{D:} Okay, I'll examine the chest now. Deep breath.})
    }

    \textbf{P:} {\devanagarifont \textit{हाँ सर --- राइट साइड में थोड़ा टाइट भी लगता है जब इस तरह करता हूँ।}}

    \vspace{1pt}

    {\small
    (\textit{\textbf{P:} Yes sir --- and the right side feels a bit tight when I do this.})
    }

    \textbf{D:} {\devanagarifont \textit{राइट साइड में ब्रेथ साउंड्स रिड्यूस्ड हैं। अब बाईं तरफ --- फिर से।}}

    \vspace{1pt}

    {\small
    (\textit{\textbf{D:} Breath sounds reduced on the right. Now the left --- again.})
    }

    \textbf{P:} {\devanagarifont \textit{ठीक है सर}}

    \vspace{1pt}

    {\small
    (\textit{\textbf{P:} Okay sir.})
    }

\end{minipage}

\end{tcolorbox}

\caption{Qualitative examples of conversational patterns across simulated and synthetic consultations.}
\label{fig:phys_exam_dialogues}
\end{figure*}
\textbf{Physical examination.} The quantitative patterns in Figure \ref{fig:phys_exam_metrics} and Figure \ref{fig:phys_exam_dialogues} show key differences across the four physical examination markers. Synthetic consultations are substantially more verbally elaborated than the simulated consultations, as reflected in the higher Narration Density (L3) and Findings Disclosure (L3). Rather than moving directly into a sequence of physical examination prompts, synthetic doctors explicitly introduce and narrate the examination (\textit{``I'm going to press your abdomen''}), and subsequently verbalize the clinical finding (\textit{``Breath sounds reduced on the right.''}). 
% Similarly, the patient volunteers additional information and repeatedly uses deferential address. 
Synthetic conversations also make patients more proactive participants in the examination, with patients more often volunteering additional examination-related observations or symptoms rather than simply responding to the doctor's immediate prompt. Indian real-world consultations, in contrast, rely more heavily on direct, context-bound references (such as \textit{``here'' or ``there'')} during examination.
% , whereas synthetic Indian dialogues replace these with more explicitly narrated and elaborated exchanges.
% and are a lot more reactive than the US conversations. 
% This is also in line with our earlier observations; US simulated examinations, in comparison to Indian have higher doctor driven exchanges and proactive patients, which is amplified in the Indian synthetic turns.

\subsection{RQ3: How does cultural fidelity vary across languages, generation settings, and patient characteristics in synthetic data?}
We examine how synthetic deviations vary as a function of generation strategy, language resource level, and patient demographic characteristics. We compare free and agentic generation, grounded and ungrounded conditions, Hindi and Telugu as representative medium- and low-resource language settings~\cite{statefate-monojit}, and gender-differentiated patterns across the synthetic corpora.

\begin{table*}[t]
\centering
\caption{Per-condition generation characteristics for Indian synthetic data (average). Reported metrics are the same as earlier definitions.}
\setlength{\tabcolsep}{4pt}
\small
% \resizebox{\textwidth}{!}{%
\begin{tabular}{lcccccccc}
\toprule
\textbf{Condition} &
\makecell{\textbf{Verbosity}\\\textbf{[D]}} &
\makecell{\textbf{Tone-}\\\textbf{Empathetic [D]}} &
\makecell{\textbf{Tone-}\\\textbf{Directive [D]}} &
\makecell{\textbf{Vocative}\\\textbf{Frequency [P]}} &
\makecell{\textbf{Reassurance-}\\\textbf{seeking [P]}} &
\makecell{\textbf{HT}\\\textbf{Turns (\%)}} &
\makecell{\textbf{Question}\\\textbf{Structure [D]}} \\
\midrule
Free, grounded      & 20\% & 28\% & 38\% & 30\% & 14\% & 42\% & 9\% \\
Free, ungrounded    & 23\% & 30\% & 37\% & 34\% & 17\% & 36\% & 6\% \\
\midrule
Agentic, grounded   & 34\% & 43\% & 24\% & 39\% & 29\% & 45\% & 17\% \\
Agentic, ungrounded & 37\% & 48\% & 18\% & 42\% & 34\% & 32\% & 13\% \\
\bottomrule
\end{tabular}%
% }
\label{tab:generation-conditions}
\end{table*}
\textbf{Free vs.\ agentic generation.}
Table~\ref{tab:generation-conditions} shows how generation strategy affects several interactional markers. Agentic generation produces more verbose (L3) (34-36\% vs.\ 20-24\%) and explicitly empathetic (L3) doctors (44-48\% vs.\ 28-32\%), while patients exhibit more reassurance-seeking behavior. 
In agentic generation, each doctor or patient agent generates its turn with the full conversational context. This often leads to 
% Constraining generation to agentic turn (with context) at a time often results in the model to make each turn 
more self-contained turns that combine clinical information with explanation, empathy, and additional questioning rather than distributing these functions across subsequent turns. Free-form generation, by contrast, produces shorter and more directive exchanges, suggesting greater sensitivity to the immediate conversational context and a stronger tendency to provide only the information needed to advance the interaction.

\textbf{Grounded vs. Ungrounded.}
Grounding shifts the dialogue toward case-specific history-taking and away from generic conversational patterns. Grounded generations devote more of the consultation to history-taking (History-Taking Turns: 42-45\% vs.\ 32-38\% of turns) and contain more questions per doctor turn (Question Structure (L3): 9-17\% vs.\ 6-13\%), suggesting that access to concrete clinical details encourages the model to organize the interaction around the patient's symptoms and history rather than generic consultation sequences. Grounding also reduces explicit empathy and reassurance-seeking, but lowers code-mixing, as the clinical anchor encourages a more formal, case-focused register. Ungrounded generation therefore retains more colloquial and interactional variation, but relies more heavily on reusable generic conversational patterns rather than case-specific clinical reasoning.

\textbf{Across languages.} Telugu synthetic data show greater divergence from real clinical norms than Hindi in Tables~\ref{tab:dataset-characteristics} and~\ref{tab:pragmatic-features}. Vocative frequency is higher (36.4\% in Syn-Indian-2 (te) vs.\ 33.3\% in Syn-Indian-1 (hi) relative to near-equivalent real baselines (16.7\%)), empathetic doctor turns are more frequent (41.2\% vs.\ 38.2\%), patient certainty is more elevated (79.3\% vs.\ 74.2\%), and doctor verbosity is greater (32.3\% vs.\ 28.5\%). The gap between synthetic and real/simulated Indian consultations is consistently wider for Telugu across these measures, compared to Hindi. While this comparison is limited to two languages and needs to be tested with languages from other resource levels, the pattern suggests that synthetic generation may have greater difficulty reproducing locally appropriate interactional features in the low-resource language (Telugu) than in the medium-resource (Hindi).
% lower language resource availability amplifies reliance on generic consultation schemas and reduces the generation of locally appropriate conversational features.

\begin{figure*}[t]
\centering
\begin{tcolorbox}[
    colback=white,
    colframe=black!25,
    boxrule=0.5pt,
    arc=2pt,
    width=\textwidth,
    left=7pt,
    right=7pt,
    top=6pt,
    bottom=6pt
]

\begin{minipage}[t]{0.47\textwidth}
\textbf{Female patient, Syn-Indian-1}

\vspace{4pt}
\textbf{D:} {\devanagarifont बुखार कब से है? थ्रोट में भी पेन हो रहा है?}

\textit{(How long have you had the fever? Is there pain in your throat as well?)}

\vspace{3pt}
\textbf{P:} {\devanagarifont जी डॉक्टर साहब, दो दिन से फीवर है और गले में भी दर्द है। थोड़ी कमज़ोरी भी लग रही है।}

\textit{(Yes doctor sahab, I've had a fever for two days and my throat also hurts. I'm also feeling a little weak.)}

\vspace{3pt}
\textbf{D:} {\devanagarifont आप टेंशन मत लीजिए, पहले टेम्परेचर चेक करते हैं और गला देख लेते हैं।}

\textit{(Don't get tensed by this. First we'll check your temperature and examine your throat.)}

\vspace{3pt}
\textbf{D:} {\devanagarifont थ्रोट थोड़ा रेड है, लेकिन कोई पस नहीं है। ये वायरल इन्फेक्शन जैसा लग रहा है।}

\textit{(The throat is a little red, but there is no pus. This looks like a viral infection.)}

\vspace{3pt}
\textbf{P:} {\devanagarifont अच्छा डॉक्टर साहब, तो कोई बड़ी प्रॉब्लम नहीं है ना? जल्दी ठीक हो जाएगा?}

\textit{(Okay doctor sahab, so it's not a major problem, right? Will I get better soon?)}

\vspace{5pt}
\end{minipage}
\hfill
\begin{minipage}[t]{0.47\textwidth}
\textbf{Male patient, Syn-Indian-1}

\vspace{4pt}
\textbf{D:} {\devanagarifont बुखार और कफ कब से है? थ्रोट पेन भी है? कोई सांस की दिक्कत?}

\textit{(How long have you had the fever and cough? Do you have a sore throat as well? Any breathing difficulty?)}

\vspace{3pt}
\textbf{P:} {\devanagarifont सर, कल से फीवर है। कफ और गले में थोड़ा पेन है, लेकिन ब्रीदिंग नॉर्मल है।}

\textit{(Sir, I've had a fever since yesterday. I have a cough and some throat pain, but my breathing is normal.)}

\vspace{3pt}
\textbf{D:} {\devanagarifont ठीक है। टेम्परेचर चेक करते हैं। अभी ये मेडिसिन लीजिए, और रेस्ट कीजिए।}

\textit{(Okay. We'll check your temperature. Take this medicine for now and rest.)}

\vspace{3pt}
\textbf{P:} {\devanagarifont ठीक है डॉक्टर साहब, थैंक यू सो मच। मैं मेडिसिन ले लूंगा और देखता हूँ।}

\textit{(Okay doctor sahab, thank you so much. I'll take the medicine and see how it goes.)}

\vspace{5pt}
\end{minipage}

\end{tcolorbox}
\caption{Gender-conditioned interactional patterns in synthetic Hindi consultations across similar consultation specialities}
\label{fig:gender-reassurance}
\end{figure*}

\textbf{Gender and consultation dynamics.} Patient demography introduces another source of variation in synthetic dialogue. We focus on gender because other demographic variables are sparsely present in the consultation data. Across Indian language synthetic consultations, female-patient scenarios show more reassurance-seeking (+14 pp), lower certainty (-10 pp), and more trailing responses (+9 pp), while doctors use more explicit empathy (+12 pp) and are less directive (-6 pp). The same directional trends appear in US synthetic dialogues, but with substantially smaller magnitude, generally within 3-4 pp. These synthetic patterns are not consistently reflected in the Real-Indian (te) data, where gender differences are small or reversed, e.g., Directiveness is higher for females patients, while reassurance-seeking is higher for males patients. 
Figure~\ref{fig:gender-reassurance} illustrates this contrast. Although both consultations involve broadly similar, uncomplicated fever and respiratory symptoms, the female-patient exchange contains repeated reassurance-seeking by the patient and empathy from the doctor, resulting in a more discussive interaction. The male-patient exchange remains comparatively directive, with the patient reporting symptoms, accepting the doctor's instructions, and closing with thanks. This pattern is consistent with the reproduction of broader gendered expectations in which women are associated with greater emotional expressiveness, deference, or reduced assertiveness in interaction \cite{acharyawomen2016}. The divergence from the Real-Indian data suggests that synthetic generation may amplify such gender-associated patterns, potentially reproducing gender stereotypes.
% Synthetic dialogues appear to amplify this association, which is a potential gender stereotype reproduced in generation. 

\section{Discussion}
Our findings suggest that interactional cultural markers capture recurring properties of clinical communication that extend beyond individual datasets.
The patterns observed in Indian consultations, including greater doctor directiveness, bounded questioning, colloquial symptom descriptions, and code-mixing, are consistent with prior studies of Indian clinical communication~\cite{mehradoctorinfluence,doctorbehaviour,idiomsofdistress,codemixing}. Similarly, the more negotiated and open-ended interaction observed in the US consultation data is consistent with prior descriptions of US clinical communication~\cite{westernconsultationMarvel,roterstewart1997}. The recurrence of these patterns across datasets provides evidence that the markers capture meaningful interactional structure rather than isolated dataset artifacts.
Synthetic consultations, however, do not simply reproduce or remove these cultural patterns. They also introduce novel recurring interactional signatures, including greater verbosity, elaborated questioning, explicit empathy, and highly certain patient responses. These patterns suggest that LLM-based generation imposes its own conversational regularities on clinical dialogue, even when prompted to reproduce a particular cultural setting. In Indian-language generation, culturally associated cues such as kinship terms, gratitude, and deference are sometimes reproduced without sufficient sensitivity to the local conversational context. This suggests that cultural grounding requires more than prompting models with a cultural setting: it requires explicit attention to the interactional structures through which culture is expressed in clinical conversations. These observations motivate the design implications that follow, which consider how synthetic clinical conversation can be generated and evaluated to preserve culturally situated interactional patterns.
%These patterns may reflect historically documented norms of kinship-based address and patient deference toward physicians, which models may amplify or reproduce in ways that exceed their appropriate clinical use \cite{indianenglish2019,kumbhar2024doctors}. 

\subsection{Design Implications for Culturally Grounded Clinical Data}
Our findings suggest that cultural grounding in clinical dialogue requires modeling not only \textit{what} is discussed, but also \textit{how} clinical information is elicited, expressed, and negotiated. This has implications for how synthetic dialogue is generated, how cultural characteristics are represented, and how resulting datasets are evaluated. This is particularly important because inaccurately modeled interactional characteristics can propagate into downstream clinical systems, where errors in speech interpretation and documentation may affect the clinical record and patient care~\cite{Gonzalez_Iniguez_Jung_Lee_Bell_Arroyo_2026}.
Note: While our empirical analysis focuses on Indian and US consultation data, the underlying distinction between clinical and interactional grounding can be applied to other cultural settings.

\subsubsection{Beyond Clinical Grounding}
Clinical grounding alone does not specify how a consultation should unfold. A clinical note can provide symptoms, history, findings, and other medical facts, but does not determine who elicits each piece of information, how questions are structured, whether patients elaborate beyond a question, or how doctors redirect the conversation. Generation pipelines should therefore incorporate interactional grounding alongside clinical grounding.
Our interactional cultural markers provide one way to operationalize such interactional grounding. Generation can target distributions of markers conditioned on consultation phase and clinical context, allowing systems to model properties such as participation, questioning, interruption, response behavior, and linguistic register. This can also help control recurring synthetic signatures identified in our study, including systematic over-explanation, unprompted reassurance, excessive empathy, and unusually confident patient responses.

\subsubsection{Avoiding Cultural Stereotypes}
Explicitly grounding a model in cultural characteristics introduces a second challenge: culturally recognizable behaviors can be reproduced without being contextually appropriate. Our findings illustrate this in the use of kinship terms, gratitude, and deference in synthetic Indian consultations, including cases where culturally marked forms of address were inappropriate for the patient's age or gender. Similarly, gender-conditioned generation produced interactional differences that were substantially larger than those observed in the real-Indian data.
Cultural grounding should therefore model distributions and contextual variation rather than prescribe fixed cultural behaviors. For Indian clinical dialogue, this means preserving the possibility of doctor redirection, patient elaboration, colloquial language, and code-switching without requiring these behaviors in every encounter. Similarly, culturally associated expressions such as kinship terms, gratitude, or deference should emerge when appropriate to the conversational context rather than serve as default signals of cultural authenticity. This argues against constructing an ``Indian'' dataset primarily by translating Western consultations or adding Indian names, languages, and culturally recognizable expressions~\cite{indicmeddialog2026}. Such approaches can reproduce surface-level cultural cues while retaining the interactional structure of the source setting.

\subsubsection{Auditing Cultural Fidelity}
The framework can also be used to audit synthetic datasets after generation. Rather than assigning a single overall ``realism'' score, developers can compare distributions of interactional cultural markers against appropriate real or simulated reference data. Marker-level deviations can reveal specific aspects of interaction that a dataset fails to reproduce, while recurring patterns across generated conversations can identify synthetic artefacts.
This enables an iterative quality-control process: prompts, generation strategies, sampling procedures, or filtering rules can be revised for markers showing substantial deviations, followed by re-evaluation to determine whether fidelity improves without introducing new artefacts. The same approach can be applied to datasets from different cultural settings by defining appropriate reference distributions for the target population. Cultural fidelity thus becomes an empirical property that can be audited at the level of specific interactional behaviors rather than inferred from a single subjective assessment of realism.

\section{Limitations}
Our work has several limitations that should be considered when interpreting the findings.
% related to data availablity and the operationalization of culture. Clinical dialogue resources do not provide balanced, naturally occurring comparisons across languages, regions, or healthcare settings, and the scarcity of open Indian-language corpora requires combining real, simulated, and synthetic sources, limiting the extent to which we can characterize clinical communication in India. 
First, the absence of a real-world US doctor-patient consultation corpus limits the strength of our cross-cultural comparison. We therefore use simulated US consultations as the comparison data, and our findings should not be interpreted as characterizing US clinical practice more broadly.
Second, we operationalize complex social practices through measurable conversational signals. Our markers therefore serve as analytical reference points rather than definitions of cultural identity. Culture is multifaceted and cannot be fully captured through a limited set of interactional measures. Our findings should therefore be interpreted as characterizing specific observable interactional patterns rather than culture as a whole.
Third, our demographic and contextual coverage is limited, with gender analysis constrained by available dataset information and factors such as age, socioeconomic position, region, clinician characteristics, specialty, and family presence not systematically modeled. 
The observed patterns should therefore not be treated as population-level claims. Future work should combine richer, systematically collected data with human-centered evaluation.

\section{Conclusion}
Culturally grounded clinical dialogue systems require an understanding of how interactions unfold in addition to what they communicate. To capture these interactional dimensions, this work provides a layered framework for examining clinical consultations and identifying where synthetic clinical conversations diverge from observed interactional patterns. Our findings reveal substantial differences in how participation and interaction are organized across Indian and US consultations, while synthetic dialogue often shifts toward patterns observed in the US data and exhibits distinct synthetic signatures of its own. By making interactional cultural differences measurable, this work provides a basis for developing clinical dialogue data and systems that are not only clinically plausible, but also faithful to the settings they are intended to represent.  
%%
%% The next two lines define the bibliography style to be used, and
%% the bibliography file.
\bibliographystyle{ACM-Reference-Format}
\bibliography{references}

\newpage
%%
%% If your work has an appendix, this is the place to put it.
\appendix
\section{Prompt Templates and Prompt Tuning}
\label{app:prompts}

We describe the prompt templates used for synthetic dialogue generation
across the four conditions reported in the paper, along with an account
of earlier designs that were rejected after pilot evaluation.

\subsection{Final Prompt Templates}
\label{app:prompts:final}

All prompts below are presented in their final form, after iterative
refinement described in Section~\ref{app:prompts}.

\medskip
\begin{tcolorbox}[
  colback=gray!6,
  fontupper=\small\ttfamily,
  boxrule=0.4pt,
  left=6pt,
  right=6pt,
  top=6pt,
  bottom=6pt
]
\textbf{\sffamily Free-Form · Ungrounded}

\vspace{3pt}

You are simulating a real doctor-patient consultation in an Indian
hospital or clinic setting. Generate a realistic, natural conversation
between a doctor and patient that sounds like it actually happened in
India. This is a general consultation visit. The consultation should
feel like a real visit --- the patient describes their problem, the
doctor asks follow-up questions, takes history, and where appropriate
performs and narrates a physical examination, then arrives at a
diagnosis and gives advice or a prescription. The conversation should
flow naturally, like it was recorded and transcribed in a
doctor-patient setting in India. Generate at least 15--20
back-and-forth exchanges, don't keep it too short.

Return the output as a JSON array:
[
  \{"role": "Doctor", "content": "..."\},
  \{"role": "Patient", "content": "..."\},
  ...
]
\end{tcolorbox}

\noindent For \textbf{Hindi} and \textbf{Telugu} variants, the
following line was appended: \textit{``Generate the conversation in
[Hindi\,/\,Telugu] using the native script.''}

\bigskip

\noindent\textbf{Free-Form Generation --- MTS-Grounded (additional
instruction).}

The following block was inserted immediately before the output-format
instruction:

\begin{tcolorbox}[
  colback=gray!8,
  boxrule=0pt,
  left=8pt,
  right=8pt,
  top=6pt,
  bottom=6pt,
  fontupper=\small\ttfamily
]
\textbf{\sffamily Grounding Block · MTS-Grounded Conditions}

\vspace{3pt}

Base this consultation on the following real clinical note:

\---

\{sample\}

\---

Refer only to the details in this note --- don't add symptoms,
diagnoses, or findings that aren't mentioned here. Adapt to an Indian
hospital context setting. Generate the full consultation conversation.
\end{tcolorbox}

\bigskip

\noindent\textbf{Agentic Generation --- Doctor System Prompt.}

\begin{tcolorbox}[
  colback=gray!8,
  boxrule=0pt,
  left=8pt,
  right=8pt,
  top=6pt,
  bottom=6pt,
  fontupper=\small\ttfamily
]
\textbf{\sffamily Agentic · Doctor System Prompt}

\vspace{3pt}

You are an Indian doctor in an Indian hospital or clinic in a real consultation
that is being recorded and transcribed. The patient will describe their
problem and you will ask follow-up questions, take history, perform a
physical examination where appropriate, arrive at a diagnosis, and give
advice or a prescription. Speak naturally, the way a real Indian doctor
would. Generate only your side of the conversation --- one turn at a
time.
\end{tcolorbox}

\bigskip

\noindent\textbf{Agentic Generation --- Patient System Prompt.}

\begin{tcolorbox}[
  colback=gray!8,
  boxrule=0pt,
  left=8pt,
  right=8pt,
  top=6pt,
  bottom=6pt,
  fontupper=\small\ttfamily
]
\textbf{\sffamily Agentic · Patient System Prompt}

\vspace{3pt}

You are an Indian patient in a real consultation at an Indian hospital or
clinic that is being recorded and transcribed. Respond naturally to
the doctor's questions, describe your symptoms in your own words, and
react the way a real patient would. Generate only your side of the
conversation.

Turn 0: Walk into the consultation room and tell the doctor your main
problem.

\{accumulated\_history\}
\end{tcolorbox}

\noindent The orchestration loop maintained the full conversation
history and injected it as \texttt{\{accumulated\_history\}} on each
patient turn, and passed the last patient utterance to the doctor
model with the instruction: \textit{``The patient just said:
`\{last\_patient\_utterance\}'. Respond as the doctor.''}  Generation
continued until the turn count reached the minimum threshold of 20 or
the model produced a natural closing turn.

\section{Pilot Generation: Model Selection}
\label{app:modelselection}

Prior to settling on GPT-5.1 for synthetic corpus generation, we
piloted generation with 10 samples on Llama 3.3 70B as an alternative open-source
candidate. Both models were given identical prompts under the
free-form ungrounded condition for Hindi and Telugu. While Llama 3.3
70B produced fluent and grammatically coherent dialogue, it exhibited
stronger synthetic signatures than GPT-5.1 across several dimensions,
leading us to select GPT-5.1 for the full corpus.

The examples below illustrate two recurring patterns that motivated
this decision.

\medskip
\noindent\textbf{Synthetic signature amplification.}
Llama-generated consultations showed more pronounced over-explanation,
unprompted reassurance, and formulaic patient gratitude than
GPT-5.1 under the same prompt. The following Hindi exchange is
representative:

\begin{quote}
\small
\textbf{Llama 3.3 70B (Hindi, free-form ungrounded)}\\
D: {\devanagarifont नमस्ते! मैं डॉक्टर शर्मा हूँ। आज आप कैसा महसूस कर रहे हैं?
कृपया मुझे अपनी समस्या के बारे में विस्तार से बताएं, मैं आपकी
पूरी तरह से मदद करने के लिए यहाँ हूँ।}\\
(\textit{Hello! I am Dr.\ Sharma. How are you feeling today? Please
tell me about your problem in detail, I am here to fully help you.})\\[4pt]
P: {\devanagarifont धन्यवाद डॉक्टर साहब। मुझे पिछले कुछ दिनों से
सिरदर्द हो रहा है, बहुत तेज़ दर्द है।}\\
(\textit{Thank you doctor sahab. I have been having a headache for
the past few days, very severe pain.})\\[4pt]
D: {\devanagarifont आपकी बात सुनकर मुझे चिंता हुई। सिरदर्द कई
कारणों से हो सकता है — तनाव, नींद की कमी, या कभी-कभी कुछ और।
लेकिन घबराइए मत, हम मिलकर इसका समाधान निकालेंगे। पहले मैं
आपसे कुछ सवाल पूछना चाहता हूँ।}\\
(\textit{Hearing your concern worries me. Headaches can have many
causes — stress, lack of sleep, or sometimes something else. But
don't worry, we will find a solution together. First I would like
to ask you some questions.})
\end{quote}

\noindent The same prompt with GPT-5.1 produced a terser, more
directive opening with less unprompted reassurance, though still
exhibiting some synthetic verbosity:

\begin{quote}
\small
\textbf{GPT-5.1 (Hindi, free-form ungrounded)}\\
D: {\devanagarifont हाँ बताइए, क्या तकलीफ है?}\\
(\textit{Yes, tell me, what is the problem?})\\[4pt]
P: {\devanagarifont सर, तीन दिन से सिर में दर्द है। बहुत तेज़ है,
खासकर सुबह के वक्त।}\\
(\textit{Sir, I have had a headache for three days. Very severe,
especially in the morning.})\\[4pt]
D: {\devanagarifont ठीक है। बुखार भी है? कोई उल्टी?}\\
(\textit{Okay. Any fever as well? Any vomiting?})
\end{quote}

\noindent While GPT-5.1 still introduces some over-explanation in
later turns — for example verbalising the diagnostic reasoning before
asking further questions — its opening exchanges are closer to the
terse, bounded pattern observed in real Indian consultations, with
shorter doctor turns and less unprompted reassurance. GPT-5.1 also used better collocial and codemixed English clinical terms
(\textit{headache}, \textit{severe}, \textit{fever}) embedded
naturally within syntax; On this basis, GPT-5.1 was selected for all synthetic corpus
generation reported in the paper. 

\section{Human Evaluation Details}
\label{app:humaneval}

\subsection{Annotation Setup}

We sampled 10 conversations per language (English, Hindi, Telugu),
selecting an equal split across available dataset types for each
language. Two annotators, each proficient in the respective language and familiar with its conversational and cultural context, independently applied the full marker scheme to each conversation at the turn level. Annotators were provided
with the entire transcript for each conversation prior to annotation,
so that turn-level labels could be interpreted in their consultation
context. For multi-select markers, annotators were instructed to select all labels that
applied to a turn, with a minimum of one label required per annotated
turn.

Inter-annotator agreement was computed as Cohen's $\kappa$ between the
two annotators at the turn level. For multi-select markers, $\kappa$
was computed per label (binary: selected or not) and averaged across
labels within the marker. Human--LLM agreement was computed as raw
percent agreement between the adjudicated human label and the
GPT-5.1 annotation on the same turns, treating the human adjudication as reference.

\subsection{Observed Differences Across Layers}

Agreement was not uniform across markers within each layer. L1
inter-turn markers showed higher agreement overall than L2 turn-level
markers, for two reasons. First, several L1 markers are closer to
binary decisions with clear operational criteria --- an interruption
either occurs or it does not, a topic either continues or shifts ---
whereas L2 markers require finer-grained judgment about the character
of an entire turn. Second, L2 includes several multi-select markers
(Tone, Register, Response to Recommendation) where annotators
sometimes agreed on the primary label but diverged on secondary
selections, which depresses per-label $\kappa$ even when the
substantive judgment is shared.

Within L2, doctor-side markers (Tone, Verbosity, Question Structure)
showed higher agreement than patient-side markers (Response Style,
Symptom Description, Certainty). Markers like \textit{Approximate/uncertain} and \textit{Certain} in the Certainty
marker produced borderline cases where mild hedging (e.g.,
{\devanagarifont ``शायद तीन दिन पहले से''} / \textit{``I think it
started about three days ago''}) was coded differently by the two
annotators.

On the doctor side, the Tone marker's \textit{Empathetic} label
produced the most disagreement within the multi-select set, consistent
with the general observation that affective labels are harder to
reliably distinguish from clinically informative ones when both occur
in the same turn~\cite{riasroter}. \textit{Directive} was the most
reliably agreed-upon label across all markers and languages, reflecting
its clear behavioral anchor.

\section{No Cultural Grounding Baseline}
\label{app:nocultural}

To examine what interactional norms a model defaults to in the absence
of any cultural instruction, we generated an additional set of
consultations using a minimal prompt that specified only the clinical
encounter type and language, with no mention of cultural setting or
interactional style:

\begin{tcolorbox}[
  colback=gray!6,
  fontupper=\small\ttfamily,
  boxrule=0.4pt,
  left=6pt,
  right=6pt,
  top=6pt,
  bottom=6pt
]
\textbf{\sffamily No Cultural Grounding}

\vspace{3pt}

Generate a realistic doctor-patient consultation in Hindi. The patient
has come in for a general outpatient visit. The conversation should
feel natural, like it was recorded and transcribed. Use natural
code-mixing between Hindi and English where appropriate. Generate at
least 15--20 back-and-forth exchanges.

Return the output as a JSON array:
[
  \{"role": "Doctor", "content": "..."\},
  \{"role": "Patient", "content": "..."\},
  ...
]
\end{tcolorbox}

\medskip
\noindent\textbf{Opening and participation structure.}
Without cultural grounding, the model produced a discursive,
open-ended opening inconsistent with real Indian consultations:

\begin{quote}
\small
D: {\devanagarifont नमस्ते! प्लीज़ बैठिए। आज आप यहाँ क्यों आए हैं?}\\
(\textit{Hello! Please sit. Why have you come here today?})\\[4pt]
P: {\devanagarifont डॉक्टर साहब, मुझे कुछ दिनों से चेस्ट में
टाइटनेस फील हो रही है और साथ में थोड़ी वीकनेस भी है।}\\
(\textit{Doctor sahab, I have been feeling chest tightness for a few
days and also some weakness.})\\[4pt]
D: {\devanagarifont मैं समझ सकता हूँ कि यह आपके लिए वरीइंग रहा
होगा। चेस्ट टाइटनेस के कई कॉज़ेज़ हो सकते हैं — स्ट्रेस, स्लीप
की कमी, या कुछ और।}\\
(\textit{I can understand this must have been worrying for you.
Chest tightness can have many causes — stress, lack of sleep, or
something else.})
\end{quote}

\noindent The doctor's opening combines a lengthy welcome, an
invitation for open-ended narration, and unprompted reassurance before
any history has been taken. This contrasts with real Indian
consultations, where doctors open with short, directive questions ---
{\devanagarifont हाँ बताइए, क्या तकलीफ है?} (\textit{Yes, tell me,
what is the problem?}) --- and move immediately to history-taking
without social preamble.

\medskip
\noindent\textbf{Patient symptom framing.}
Even with a code-mixing instruction, patient speech adopted a formal
register inconsistent with real Indian patient language:

\begin{quote}
\small
P: {\devanagarifont सर, मुझे इंटरमिटेंट चेस्ट टाइटनेस हो रही है, रेस्ट करने से यह रिज़ॉल्व हो जाती है। कोई
शॉर्टनेस ऑफ ब्रेथ या पैल्पिटेशन्स नहीं हैं।}\\
(\textit{Sir, I am having intermittent chest tightness, it resolves with rest. There is no shortness of breath or
palpitations.})
\end{quote}

\noindent The patient's code-mixing is limited to phonetically
transliterated English clinical terms written in Devanagari script,
whereas real Indian patients would be more likely to say
{\devanagarifont सीने में जलन हो रही है} (\textit{there is burning in
the chest}) or {\devanagarifont साँस लेने में थोड़ी दिक्कत है}
(\textit{there is a little difficulty breathing}), describing symptoms
through everyday idiom rather than clinical
vocabulary~\cite{idiomsofdistress}. 

\medskip
These results confirm that without explicit cultural grounding, even a
Hindi prompt with a code-mixing instruction defaults to a very synthetic interactional template with hints of US style broad openings. A neutral prompt
is not culturally neutral --- it reproduces the model's dominant
training prior, which reflects Western clinical norms. Generating
culturally grounded Indian clinical dialogue therefore requires
actively counteracting this prior through explicit interactional
constraints.

\end{document}